\documentclass[twocolumn,preprintnumbers,showkeys]{revtex4}
\usepackage{xcolor}
\usepackage{graphicx}
\usepackage{graphicx,epstopdf}
\usepackage{dcolumn}
\usepackage{bm}
\usepackage{amsmath,mathrsfs}
\usepackage{amsthm,amssymb,amsmath,mathrsfs,mathtools}
\usepackage[mathscr]{euscript}
\usepackage[colorlinks=true,citecolor=blue]{hyperref}

\begin{document}

\title{Dynamics on higher-order networks: A review}

\author{Soumen Majhi}
\email{soumen.majhi91@gmail.com}
\affiliation{Department of Mathematics, Bar-Ilan University, Ramat-Gan, 5290002, Israel}
\author{Matja{\v z} Perc}
\email{matjaz.perc@gmail.com}
\affiliation{Faculty of Natural Sciences and Mathematics, University of Maribor, Koro{\v s}ka cesta 160, 2000 Maribor, Slovenia}
\affiliation{Department of Medical Research, China Medical University Hospital, China Medical University, 404332 Taichung, Taiwan}
\affiliation{Complexity Science Hub Vienna, Josefst\"{o}dter Str. 39, 1080 Vienna, Austria}
\affiliation{Alma Mater Europaea, Slovenska ulica 17, 2000 Maribor, Slovenia}
\author{Dibakar Ghosh}
\email{dibakar@isical.ac.in}
\affiliation{Physics and Applied Mathematics Unit, Indian Statistical Institute, 203 B. T. Road, Kolkata 700108, India}

\begin{abstract}
Network science has evolved into an indispensable platform for studying complex systems. But recent research has identified limits of classical networks, where links connect pairs of nodes, to comprehensively describe group interactions. Higher-order networks, where a link can connect more than two nodes, have therefore emerged as a new frontier in network science. Since group interactions are common in social, biological, and technological systems, higher-order networks have recently led to important new discoveries across many fields of research. We here review these works, focusing in particular on the novel aspects of the dynamics that emerges on higher-order networks. We cover a variety of dynamical processes that have thus far been studied, including different synchronization phenomena, contagion processes, the evolution of cooperation, and consensus formation. We also outline open challenges and promising directions for future research.
\end{abstract}


\maketitle


\section{Introduction}

The theory of complex networks~\cite{albert2002statistical,newman2003structure} provides us with a framework for investigating the structure and dynamics of interacting systems. Indeed, network science has proven highly efficient in elucidating the dynamics of complex systems arising from many different contexts in the physical, biological as well as technological and social sciences~\cite{boccaletti2006complex,barrat2008dynamical}. Many key developments have been made in view of identifying and improving the concepts of association among the constituents of a network. To illustrate, the necessity of considering the links of networks that are different in nature has led to the formulation and detailed analysis of multilayer networks~\cite{boccaletti2014structure}. Further, time-varying networks~\cite{holme2012temporal,ghosh2022synchronized} are investigated in which interactions do not persist for all the course of time, rather come up or vanish over time. It is unquestionably true that all these developments have helped us to perceive many scenarios better, but specifically assumed dyadic or pairwise interactions as the backbone for connections among the units of the system. However, a complete explanation of many complex systems requires more to furnish. Indeed, real-world systems often exhibit higher-order relationships beyond dyadic interactions~\cite{battiston2020networks,bianconi2021higher}. For instance, group interactions take place predominantly in systems arising in neurobiology~\cite{giusti2016two,petri2014homological} and social systems~\cite{benson2016higher,bianconi2021higher} to ecology~\cite{levine2017beyond,grilli2017higher}. Network framework has been intrinsically limited to explanation through the pairwise interactions which are sufficient only to model the dyadic relationships and so larger group interactions do need a better formulation for networked systems~\cite{lambiotte2019networks,johnson2013hypernetworks,bick2021higher}. It has been argued that higher-order structures, namely hypernetworks and simplicial complexes, are excellent frameworks to characterize the organization of many social, biological and other scenarios encoded in group interactions of three or more constituents~\cite{battiston2020networks,battiston2021physics}.

\par Heretofore, not much attention has been paid to the analysis of networks exposed to higher-order interactions. However, a significant amount of recent advances is demonstrating that the incorporation of higher-order architecture can remarkably improve our understanding and prediction ability of their dynamics. The studies related to these higher-order structures have thus come to the forefront of network dynamics research. Among some highly significant researches on higher-order networks, the one by Benson et al.~\cite{benson2018simplicial} is particularly noteworthy that investigates datasets from different disciplines ranging from various social networks to biology, and demonstrates variety of characteristics of the higher-order structures emerging therein. Besides, the problem of higher-order link prediction is formulated which has been found to be essentially different from the traditional dyadic link prediction~\cite{liben2007link}. This issue of link prediction in networks subject to the presence of higher-order structures is also studied in the Ref.~\cite{sharma2021higher} while dealing with different link prediction algorithms. The inverse problem of inferring higher-order interactions from observational data has also been treated in the Ref.~\cite{beentjes2020higher}, whereas higher-order interactions are inferred from the traditional dyadic interaction network data through a Bayesian approach~\cite{young2021hypergraph}. An analytical treatment (Statistically Validated Hypergraphs) is propounded~\cite{musciotto2021detecting} for the problem of finding the most important relationships among the constituents of a higher-order network. Tie strengths are quantified considering higher-order interactions encoded by groups of three or more individuals in social networks by the measure named as `Edge PageRank'~\cite{sarker2021higher}. This measure is proved to be much more efficacious than the traditional approaches for detection of tie strength. Vector centrality measure is proposed for higher-order networks with an aim to identify the most influential nodes in the system~\cite{kovalenko2021vector}.

\par Different models of higher-order networks~\cite{bobrowski2021random} have been developed so far. Detailed analysis of models of growing simplicial complexes~\cite{kovalenko2021growing,courtney2017weighted,courtney2018dense} is presented, built upon the concept of `network geometry with flavor (NGF)'~\cite{bianconi2016network,mulder2018network}. The models yield exponential or scale-free generalized degree distribution based on the non-preferential or preferential attachment rules. `Simplicial activity driven model'~\cite{petri2018simplicial} is proposed and analyzed, that captures both the higher-order structure and the temporal nature of interactions among the nodes. `Simplicial configuration model'~\cite{young2017construction,courtney2016generalized} with an uniform Markov chain Monte Carlo sampler is introduced, even for arbitrary degree and size distributions~\cite{yen2021construction}. In order to provide formalism for modeling heterogeneous, polyadic network data, the configuration models of random hypernetworks~\cite{chodrow2020configuration} and the annotated hypergraph model~\cite{chodrow2020annotated} are presented as a generalization of directed graphs. On the other hand, higher-order network setups are utilized in order to generalize the formalism of structural controllability to time-varying networks~\cite{zhang2021higher}, for which both synthetic and real-world data sets are examined to illustrate the minimum time required to control the concerned systems.  Group research collaborations of three or more individuals are illustrated through higher-order interaction framework and further encoded under multilayer formalism with collaboration data taken from different research disciplines~\cite{vasilyeva2021multilayer}. Further, heterogeneous dynamics of higher-order structures in time-varying social networks is examined for a number of social datasets~\cite{cencetti2021temporal}.

\par The perception of higher-order clusters being particularly important, the concept of higher-order clustering coefficients are introduced in the Ref.~\cite{yin2018higher}, which quantifies the closure probability of higher-order cliques. This measure is used to examine the clustering behavior of both model and real-world networks. The problem of clustering in hypernetworks with categorical edge labels can be addressed with a procedure based upon combinatorial objective function~\cite{amburg2020clustering}. The efficiency of this mechanism is validated for edge-label community detection and clustering with time-stamped data. Simplicial communities are detected from real-world data of social networks while showing that the spectra of the Hodge Laplacian encodes the communities~\cite{krishnagopal2021spectral}. A stochastic generative model is introduced to hypernetwork clustering with heterogeneous node degree and hyperlink size distribution~\cite{chodrow2021generative}, which is shown to be highly scalable and efficient with the utilization of synthetic and various real-world data. Tudisco et al.~\cite{tudisco2021node} has recently come up with their study of a family of spectral centrality measures in order to recognize important nodes and hyperlinks in hypernetworks, which extends the existing concepts for dyadic interactions. However, the formalism constructed by Veldt et al.~\cite{veldt2021higher} using hypernetworks to measure `homophily', unravels that homophilous group configurations are impossible due to a combinatorial impossibility of hypernetworks.

\par Concerning dynamical processes, framework for hypernetwork robustness and analysis of higher-order percolation process~\cite{sun2021higher,bianconi2019percolation,lee2021homological} are put forward for multiplex hypernetworks. Analogous to the largest eigenvalue of the matrix representing interaction structure of a network built upon dyadic connections, the concept of `expansion eigenvalue' for hypernetwork dynamical processes is proposed and approximated through a mean-field approach~\cite{landry2021hypergraph}. Quite interestingly, in case of random interactions in ecological communities, the presence of higher-order species interactions can verily alter the traditional relation between diversity and stability~\cite{bairey2016high}. For instance, even though dyadic interactions cause sensitivity to the species addition, four-way interactions result in sensitivity to the removal of species. Also, the merger of the dyadic and higher-order interaction induces both upper and lower bounds on the number of species. Moreover, there exists evidence of higher-order interactions stabilizing the dynamics in ecological communities~\cite{grilli2017higher} where interaction between species is influenced by other species. In both open and closed ecological communities, higher-order interactions have noticeable impacts that stabilize the dynamics for competitive models. Stochastic models under higher-order interactions help further in sustained coexistence of species (see also the review work~\cite{levine2017beyond} and the references therein).\\

\par In the next section (Sec.~\ref{basics}), we briefly recall the basic definition of relevant terminology in higher-order interactions. We then start discussing about the phenomenon of synchronization emerging in higher-order networked systems (cf. Sec.~\ref{sync}). Then, in Sec.~\ref{soc}, we move on to explore various social dynamics evolving over higher-order structures. Specifically, we investigate the processes of contagion dynamics, consensus formation and evolutionary game dynamics in Secs.~\ref{contg},~\ref{cons} and~\ref{egd} respectively. Section~\ref{rand} deals with the dynamics of random walk and diffusion. Finally, Sec.~\ref{summ} offers the summary and discussion about potential research in the future.

\clearpage

\section{Basic Concepts}\label{basics}

\begin{figure}[ht]
	\centerline{\includegraphics[width = 6cm]{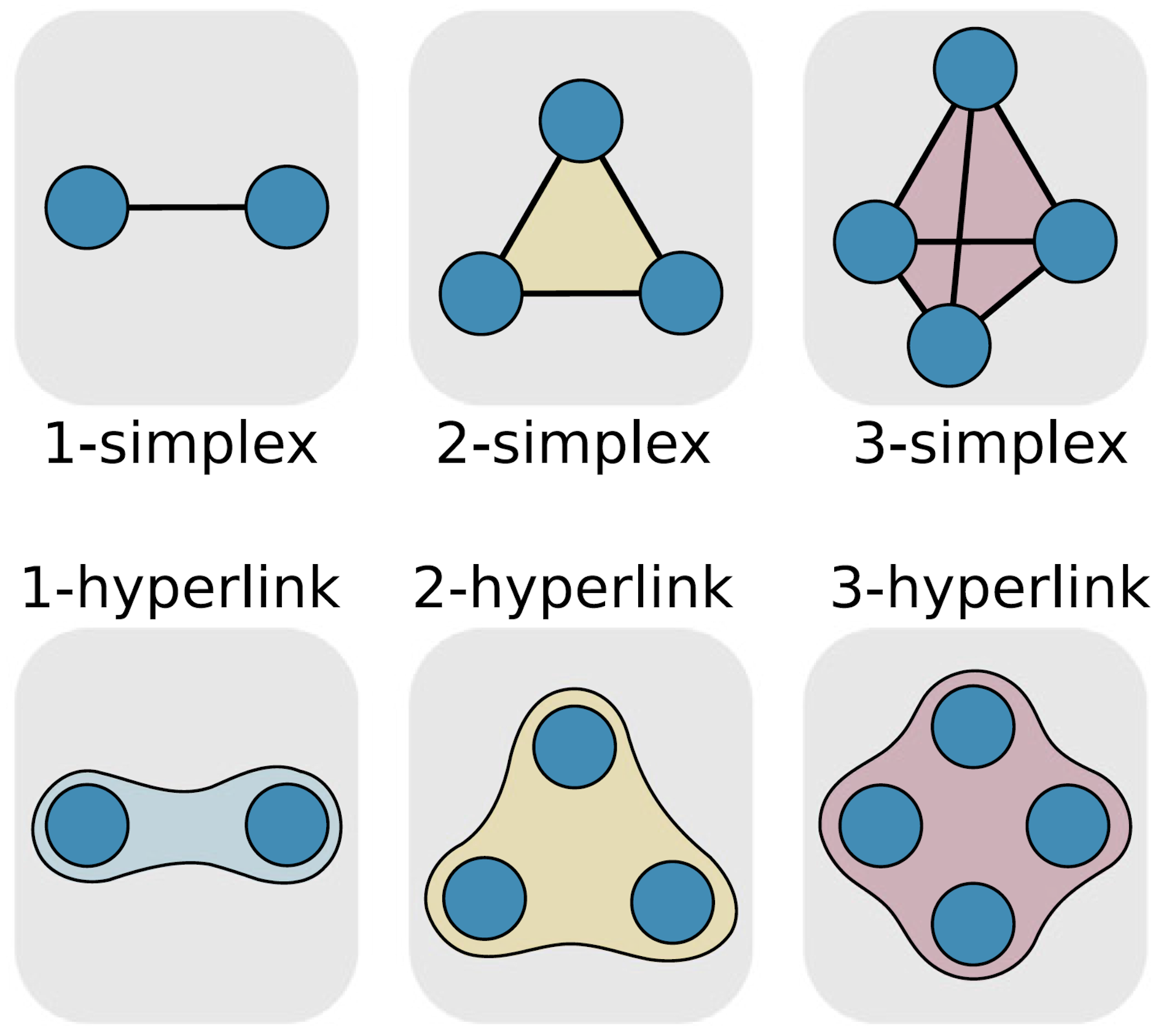}}
	\caption{The higher-order building blocks, namely the simplices (upper panel) and the hyperlinks (lower panel) of dimension $1$, $2$ and $3$. Figure reproduced from the Ref.~\cite{battiston2020networks}.}
	\label{scheme}
\end{figure}

\textit{Hyperlink}: Hyperlink is the fundamental backbone of a higher-order network, which instead of joining only two nodes (for the traditional networks of pairwise interactions), can connect any number of nodes.   \\
\textit{Hypernetwork}: Hypernetwork is a generalization of the notion of network, and is composed of hyperlinks. This implies that a hypernetwork $\mathcal{H}$ can be considered as the pair $(V,E)$ in which $V$ is the set of nodes and $E$ (a subset of the power set $P(V)$ of $V$) is the set of hyperlinks. \\
\textit{Simplex}: A $d$-dimensional simplex (or a $d$-simplex) is simply a set of $d+1$ fully interacting nodes. Essentially, a $0$-simplex is a node, a $1$-simplex is a link, a $2$-simplex is a triangle, a $3$-simplex is a tetrahedron and so on. \\
\textit{Simplicial complex}: Similar to the networks as a collection of links, a simplicial complex comprises of simplices. From the context of hypernetworks, a simplicial complex is a particular type of hypernetwork that accommodates each subset of all the hyperlinks. This means that a simplicial complex $\mathcal{S}$ is a hypernetwork which fulfills the criteria that for each $e \in E$ and $\forall e' \subseteq e~(e'\neq\emptyset)$, one also has $e' \in E$.  \\
In Fig.~\ref{scheme}, examples of simplices (in upper panel) and hyperlinks (in lower panel) of dimension from $1$ to $3$ are depicted that clarify about the higher-order building blocks based upon which the higher-order networks are generally built.

\par In the next, we focus on elaborating the principal findings and the novel effects in the dynamical processes that the higher-order interactions bring about, and hence can be of interest for the perception of a number of natural occurrences. We, however, mention here that we do not really distinguish between the dynamics on hypernetworks and dynamics on simplicial complexes, rather we present an excerpt of diverse dynamics on top of higher-order networks, in general.

\section{Synchronization}\label{sync}

The phenomenon of synchronization~\cite{pikovsky2003synchronization,boccaletti2002synchronization,wang2009synchronization,majhi2019emergence,arenas2008synchronization,majhi2017synchronization} corresponds to a process wherein interacting dynamical systems adjust certain properties of their motion to a common dynamics, and this interaction pattern plays decisive roles for the emergence of synchrony. Synchronization is considered to be one of the most important phenomenon in complex dynamical network theory, having crucial applications in several physical, biological and technological systems. Hence, there has been a strong urge in exploring different aspects of synchronization in coupled systems in the last two decades. However, only recently the investigation tends towards higher-order interactions.

\par  Specifically, three-body interactions in an ensemble of phase oscillators can give rise to infinite number of multistable synchronized attractors beyond a critical interaction strength~\cite{tanaka2011multistable}. Simplicial complexes of large ensembles of interacting oscillators are considered while modeling the three-way interactions on top of a multilayer framework~\cite{skardal2019abrupt}. A continuum of abrupt transitions to desynchronization is observed therein as a result of multistability comprising of infinite number of stable partially synchronized states. An analytical treatment is provided based upon dimensionality reduction through a variation of the Ott-Antonsen ansatz. Besides, synergistic effects of higher-order interactions of different order (namely, $1$-, $2$- and $3$-simplex) on synchrony arising in heterogeneous Kuramoto phase oscillators is analyzed~\cite{skardal2020higher}. Here, it has been shown that the interplay of these simplicial structures can yield abrupt transitions to both desynchronization and synchronization, and can stabilize strong synchrony even under repulsive pairwise interaction. The authors also enunciate the phenomena while dealing with UK power grid and Macaque brain networks. Gambuzza et al.~\cite{gambuzza2021stability} have recently presented a general framework for the stability of synchronization in networks subject to higher-order interactions of any order. The authors demonstrate the existence of complete synchrony as an invariant solution and provided the necessary conditions for the synchronous solution. The generality of the proposed formalism is elucidated considering paradigmatic chaotic R\"{o}ssler system and model systems pertinent to neurodynamics. Besides, cluster synchronization is studied in a model of simplicial complex of R\"{o}ssler oscillators.

\par Analysis of $D~(\ge 2)$-dimensional Kuramoto dynamics on top of simplicial structures ($1$- and $2$-simplices predominantly) is presented in the Ref.~\cite{dai2021d}. Theoretical analysis together with extensive numerical simulations are put forward wherein reasoning behind different synchronization patterns resulting from odd and even dimensions is explained \cite{chandra2019continuous}. Interestingly, discontinuous transition to desynchronization for any dimension at positive interaction strength, discontinuous transition for odd dimensions at zero coupling strength, and the state of partial synchronization for all odd $D$ (along with $D=2$) at negative interaction strength are witnessed. Further, a globally coupled ensemble of the $D$-dimensional Kuramoto oscillators consisting of only contrarians possesses collective synchrony in absence of any conformists, if the underlying connection among the units goes beyond dyadic interactions~\cite{kovalenko2021contrarians}. This result, in particular, is forbidden in networks with only pairwise communications. A notably interesting formulation of the higher-order Kuramoto dynamics that designates interactions among oscillators placed not only on the nodes but also on the higher-dimensional simplices like links, triangles etc. permits one to describe a topologically projected dynamics on lower- and higher-dimensional faces~\cite{millan2020explosive}. It has been shown that besides a simple continuous transition to synchronization, with an adaptive coupling between the dynamics projected on the lower- and higher- dynamical phases the networked system exhibits \textit{explosive} transition to synchrony.

\par Furthermore, the interaction between dynamical signals defined on nodes and links yields explosive topological synchronization wherein the phases ascribed on the nodes synchronize to the phases defined on the links at a discontinuous transition~\cite{ghorbanchian2021higher}. Detailed analytical treatment is provided exploring this scenario
and the associated closed hysteresis loop in the limit of large size of the networks. Besides dealing with simplicial complexes, the model has been tested on the human and \textit{C. elegans} connectomes. To be precise, a simplicial complex of $N_{[n]}$ simplices of dimension $n$ (i.e., $N_{[0]}$ nodes, $N_{[1]}$ number of links, $N_{[2]}$ triangles etc.) with ${\boldsymbol{B}}_{[n]}$ as the $n$-th incidence matrix for the $n$-th boundary operator, are assumed. Then the phase vector $\boldsymbol{\theta}=(\theta_1,\theta_2,\cdots,\theta_{N_{[0]}})$ associated to the nodes obeys the following dynamical equation:

 \begin{equation}
 	\begin{array}{lcl}\label{eq1q}
 		
 		\dot {\boldsymbol{\theta}}= \boldsymbol{\omega}-\sigma {\boldsymbol{B}}_{[1]}\mbox{sin}\Big({{\boldsymbol{B}}^T_{[1]}}\boldsymbol{\theta}\Big), 		
 	\end{array}
 \end{equation}
where $\sigma$ is the interaction strength, with each $\omega_k$ chosen from a given distribution, say a normal distribution $\omega_i \sim \mathcal{N}(\Omega_0,1/\tau_{0})$. Consequently, the associated order parameter can be written as

 \begin{equation}
	\begin{array}{lcl}\label{eq1q1}
		
		R_0= \dfrac{1}{N_{[0]}}\Big|\sum\limits_{k=1}^{N_{[0]}}e^{i\theta_k}\Big|. 		
	\end{array}
\end{equation}
Higher-order topological Kuramoto model defined on phases $\boldsymbol{\phi}=(\phi_{l_1},\phi_{l_2},\dots,\phi_{l_{N_{[1]}}})$   associated to the links is described as

\begin{equation}
	\begin{array}{lcl}\label{eq1q2}
		
		\dot {\boldsymbol{\phi}}= \boldsymbol{\tilde{\omega}}-\sigma {{\boldsymbol{B}}^T_{[1]}}\mbox{sin}\Big({{\boldsymbol{B}}_{[1]}}\boldsymbol{\phi}\Big)-\sigma {{\boldsymbol{B}}_{[2]}}\mbox{sin}\Big({{\boldsymbol{B}}^T_{[2]}}\boldsymbol{\phi}\Big), 		
	\end{array}
\end{equation}
with ${\tilde{\omega}}_l \sim \mathcal{N}(\Omega_1,1/\tau_{1})$ as the internal frequencies for the links. With the synchronization dynamics defined on the higher-order $n(=1)$-dimensional signals, the projections ${\boldsymbol{\phi}}^{[-]}$ and ${\boldsymbol{\phi}}^{[+]}$ on the $n-1$ simplices (i.e., nodes) and $n-2$ simplices (i.e., triangles) are ${\boldsymbol{\phi}}^{[-]}={{\boldsymbol{B}}_{[1]}}{\boldsymbol{\phi}}$ and ${\boldsymbol{\phi}}^{[+]}={{\boldsymbol{B}}^T_{[2]}}{\boldsymbol{\phi}}$, which respectively follow the following dynamics

\begin{equation}
	\begin{array}{lcl}\label{eq1q3}
		
		\dot {{\boldsymbol{\phi}}}^{[-]}= {{\boldsymbol{B}}_{[1]}}\boldsymbol{\tilde{\omega}}-\sigma{{\boldsymbol{L}}_{[0]}} \mbox{sin}({{\phi}}^{[-]}), \\
		\dot {{\boldsymbol{\phi}}}^{[+]}= {{\boldsymbol{B}}^T_{[2]}}\boldsymbol{\tilde{\omega}}-\sigma{{\boldsymbol{L}}^{\mbox{\tiny down}}_{[2]}} \mbox{sin}({{\phi}}^{[+]}),		
	\end{array}
\end{equation}
where ${{\boldsymbol{L}}_{[0]}}={{\boldsymbol{B}}_{[1]}}{{\boldsymbol{B}}^T_{[1]}}$ and ${{\boldsymbol{L}}^{\mbox{\tiny down}}_{[2]}}={{\boldsymbol{B}}^T_{[2]}}{{\boldsymbol{B}}_{[2]}}$, and the corresponding order parameters obtain the forms:
 \begin{equation}
	\begin{array}{lcl}\label{eq1q4}
		
		{R}^{\mbox{\tiny down}}_1= \dfrac{1}{N_{[0]}}\Big|\sum\limits_{k=1}^{N_{[0]}}e^{i{{\phi}}^{[-]}_k}\Big|,\\
~~~	{R}^{\mbox{\tiny up}}_1=	 \dfrac{1}{N_{[2]}}\Big|\sum\limits_{k=1}^{N_{[2]}}e^{i{{\phi}}^{[+]}_k}\Big|.
	\end{array}
\end{equation}
Then, unlike the adaptive coupling between these two dynamics of the same dimension as in the Ref.~\cite{millan2020explosive}, here signals of different dimensions are coupled through the order parameters of the node and link dynamics (i.e., Eqs.~\eqref{eq1q1} and~\eqref{eq1q4}). In particular, two models named as Nodes-Links (NL) and Nodes-Links-Triangles (NLT) are considered. The former one is described as

\begin{equation}
	\begin{array}{lcl}\label{eq1q5}
		\dot {\boldsymbol{\theta}}= \boldsymbol{\omega}-\sigma{R}^{\mbox{\tiny down}}_1 {\boldsymbol{B}}_{[1]}\mbox{sin}\Big({{\boldsymbol{B}}^T_{[1]}}\boldsymbol{\theta}\Big),\\
		
		\dot {\boldsymbol{\phi}}= \boldsymbol{\tilde{\omega}}-\sigma R_0 {{\boldsymbol{B}}^T_{[1]}}\mbox{sin}\Big({{\boldsymbol{B}}_{[1]}}\boldsymbol{\phi}\Big)-\sigma {{\boldsymbol{B}}_{[2]}}\mbox{sin}\Big({{\boldsymbol{B}}^T_{[2]}}\boldsymbol{\phi}\Big), 		
	\end{array}
\end{equation}
with the projected dynamics following
\begin{equation}
	\begin{array}{lcl}\label{eq1q6}
		
		\dot {{\boldsymbol{\phi}}}^{[-]}= {{\boldsymbol{B}}_{[1]}}\boldsymbol{\tilde{\omega}}-\sigma R_0{{\boldsymbol{L}}_{[0]}} \mbox{sin}({{\phi}}^{[-]}), \\
		\dot {{\boldsymbol{\phi}}}^{[+]}= {{\boldsymbol{B}}^T_{[2]}}\boldsymbol{\tilde{\omega}}-\sigma{{\boldsymbol{L}}^{\mbox{\tiny down}}_{[2]}} \mbox{sin}({{\phi}}^{[+]}).	
	\end{array}
\end{equation}
The latter one is defined as
\begin{equation}
	\begin{array}{lcl}\label{eq1q7}
		\dot {\boldsymbol{\theta}}= \boldsymbol{\omega}-\sigma{R}^{\mbox{\tiny down}}_1 {\boldsymbol{B}}_{[1]}\mbox{sin}\Big({{\boldsymbol{B}}^T_{[1]}}\boldsymbol{\theta}\Big),\\
		
		\dot {\boldsymbol{\phi}}= \boldsymbol{\tilde{\omega}}-\sigma R_0{R}^{\mbox{\tiny up}}_1 {{\boldsymbol{B}}^T_{[1]}}\mbox{sin}\Big({{\boldsymbol{B}}_{[1]}}\boldsymbol{\phi}\Big)-\sigma{R}^{\mbox{\tiny down}}_1 {{\boldsymbol{B}}_{[2]}}\mbox{sin}\Big({{\boldsymbol{B}}^T_{[2]}}\boldsymbol{\phi}\Big), 		 
	\end{array}
\end{equation}
where the projected dynamics obeys
\begin{equation}
	\begin{array}{lcl}\label{eq1q8}
		
		\dot {{\boldsymbol{\phi}}}^{[-]}= {{\boldsymbol{B}}_{[1]}}\boldsymbol{\tilde{\omega}}-\sigma R_0{R}^{\mbox{\tiny up}}_1{{\boldsymbol{L}}_{[0]}} \mbox{sin}({{\phi}}^{[-]}), \\
		\dot {{\boldsymbol{\phi}}}^{[+]}= {{\boldsymbol{B}}^T_{[2]}}\boldsymbol{\tilde{\omega}}-\sigma{R}^{\mbox{\tiny down}}_1{{\boldsymbol{L}}^{\mbox{\tiny down}}_{[2]}} \mbox{sin}({{\phi}}^{[+]}).	
	\end{array}
\end{equation}

\begin{figure*}[ht]
	\centerline{\includegraphics[width = 12cm]{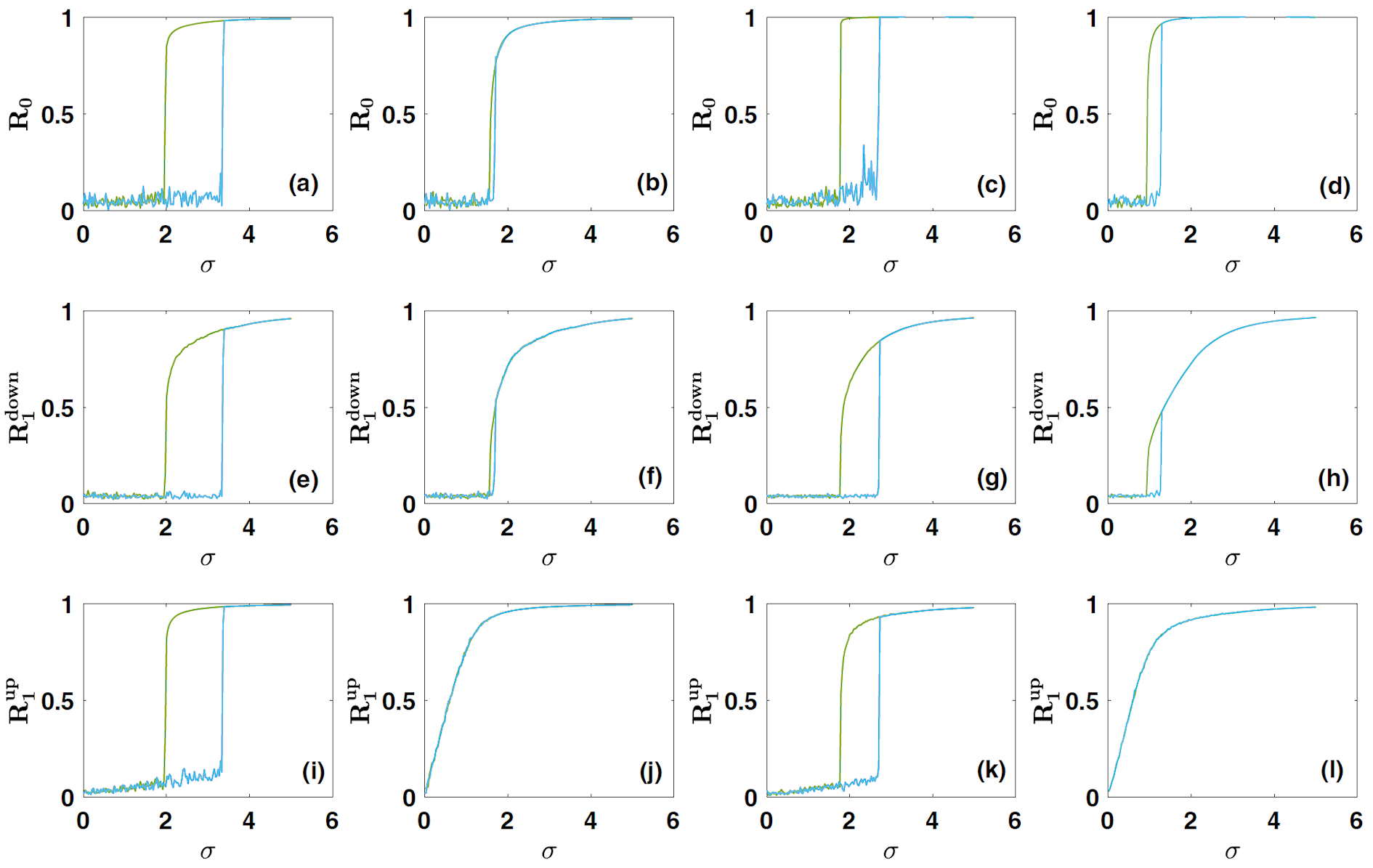}}
	\caption{ The evolution of the order parameters $R_0$, ${R}^{\mbox{\tiny down}}_1$ and ${R}^{\mbox{\tiny up}}_1$ with respect to the coupling strength $\sigma$. The first two columns (from left) correspond to the NGF model, with the last two columns representing the configuration model. Also, the second and fourth columns (from left) represent the Nodes-Links (NL) model whereas the first and third columns (from left) correspond to the Nodes-Links-Triangles (NLT) model. The cyan lines designate the forward transitions and the green lines indicate the backward transitions. For both the network models, $N_{[0]}=500$. Further, $\Omega_0=\Omega_1=2$ and $\tau_0=\tau_1=1$ are chosen. Figure reproduced from the Ref.~\cite{ghorbanchian2021higher}.}			
	\label{synch}
\end{figure*}

\par Now, with these two dynamical models NL and NLT in hand, extensive numerical simulations are carried out on two models of simplicial complexes, namely the configuration model \cite{courtney2016generalized} and the NGF model \cite{bianconi2016network}. Figure~\ref{synch} displays the values of the order parameters $R_0$, ${R}^{\mbox{\tiny down}}_1$ and ${R}^{\mbox{\tiny up}}_1$ as functions of the coupling strength $\sigma$ in top, middle and bottom panels, respectively. The first two columns correspond to the NGF model for flavor $s=-1$ and $d=3$-dimensional simplicial complexes with an underlying power-law network (exponent $\gamma=3$), whereas the last two columns are for the configuration model for power-law (exponent $\gamma=2.8$) generalized degree distribution, with $N_{[0]}=500$ nodes for both the cases. Moreover, the first and third (second and fourth) columns depict the outcomes for the NLT (NL) model. As confirmed, in both the scenarios, explosive transitions to the state of synchronization take place. The transitions occurs along with the hysteresis loop formed by the forward and backward transitions. For the NLT model, all the order parameters $R_0$, ${R}^{\mbox{\tiny up}}_1$ and ${R}^{\mbox{\tiny down}}_1$ show discontinuous transition to synchrony at the same coupling strength. But in case of the NL model, although $R_0$ and ${R}^{\mbox{\tiny down}}_1$ demonstrate discontinuous transition for some critical interaction strength, ${R}^{\mbox{\tiny up}}_1$ ensues an independent transition at the zero coupling strength for both the network models. This is because in the NL model the adaptive interaction couples only the phases ${{\boldsymbol{\phi}}}^{[-]}$ and ${\boldsymbol{\theta}}$, and not the phase ${{\boldsymbol{\phi}}}^{[+]}$. For the further details of the nice analytical treatment alongside the numerical results, please see the Ref.~\cite{ghorbanchian2021higher}.

\par Synchronization in an ensemble of Kuramoto phase oscillators subject to the interplay of interactions built upon $1$-simplex (i.e., the links) and the $2$-simplex (i.e., the triangles) faces of homogeneous $4$-dimensional simplicial complexes is reported~\cite{chutani2021hysteresis}. In presence of only dyadic interactions, increasing positive coupling strength leads to a continuous transition to complete synchrony whereas negative coupling results in a partial synchronization. It needs to be mentioned here that no synchrony is observed for negative coupling in scale-free networks. Moreover, introduction of the higher-order ($2$-simplex) coupling impedes the synchrony induced by the pairwise interaction, and causes the hysteresis loop with abrupt transition to desynchronization for negative pairwise coupling. Also, in a very recent work~\cite{kachhvah2021hebbian}, the authors assume an adaptive higher-order (triadic) interaction formalism relying on the Hebbian learning mechanism in networks of Kuramoto phase oscillators and showed that such a coupling can induce first-order transition to desynchronization. The presented scenario is further explained through a detailed mean-field analysis. Partial loss of synchronization can also be witnessed in a generalized Sakaguchi-Kuramoto model formed through the inclusion of linear and nonlinear frustrations into the simplicial Kuramoto model~\cite{arnaudon2021connecting} and weights on the simplices. A precise mathematical framework is presented in the article apart from the computational results for this frustrated model on links while dealing with measures like the order parameter and Hodge decomposition.

\par Stability of synchronization in ensembles of oscillators subject to higher-order interactions of any order and built upon any complex underlying hypernetwork structure with arbitrary coupling functions can be analyzed through a general formalism of \textit{multiorder Laplacian}~\cite{lucas2020multiorder}. Different network setups are investigated including the one that deals with both attractive and repulsive interactions~\cite{majhi2020perspective}.  From synthetic to empirical datasets are studied under this framework. Higher-order interactions embedded in clique complexes can optimize the collective synchronization in Kuramoto model subject to equitable increase in the strength of the higher-order connections relative to the pairwise interactions~\cite{skardal2021higher}. Synchronization dynamics and formation of Turing patterns in nonlinear chaotic systems are studied in higher-order networks while opting the master stability function (MSF) framework and analyzing an appropriate combinatorial Laplacian~\cite{carletti2020dynamical}. The processes are examined for general hypernetworks with heterogeneous distribution of the hyperlinks, and not restricting to any specific forms of the coupling function. Different synchronization patterns from cluster synchrony to chimeras are realized in generalized networks, including multilayer networks, hypernetworks and time-varying networks through the simultaneous block diagonalization (SBD) approach~\cite{zhang2021unified}. MSF formalism has also been generalized for hypernetworks in the Ref.~\cite{mulas2020coupled} to treat the linear stability of the phenomenon of synchronization, where the special class of Laplace-type interactions have also been examined. The dynamical systems known as coupled map lattices are extended to the scenario of higher-order interaction, namely to the concept of coupled hypergraph maps~\cite{bohle2021coupled}. The process of cluster synchronization is investigated in such a system through the analysis of hypernetwork Laplacian for different chaotic discrete dynamical systems. Very recently, in the Ref. \cite{parastesh2022synchronization} the authors consider three-body interactions along with the dyadic couplings for interacting Hindmarsh-Rose neuronal model while deriving the necessary conditions for the emergence of synchrony by means of linear stability analysis.

\section{Social dynamics}\label{soc}

Diverse social processes have always been a major area of research in complexity science. A wide list of scenarios ranging from opinion, cultural and language dynamics to crowd behavior, hierarchy formation, human dynamics, evolution of cooperation and social spreading is considerably influenced by peer-to-peer interaction among individuals embedded in social networks. Such contagion effects directs researchers to explore the dynamics from mathematical point of view. For this purpose, network science has emerged to play the most significant role. In the past decade, we have witnessed a golden age in the study of social dynamics over networks, from different perspectives. The research communities have long been interested in the interactions between individuals leading to diverse emerging behavior. However, as noted in the review by Castellano et al.~\cite{castellano2009statistical}, and the Ref.~\cite{malmgren2009universality}, there are significant aspects of real social contagion phenomena that needs to be captured with much more sophisticated approaches than before, from the perspective of both dynamics on networks and dynamics of networks. In the following, we go through different genres of studies of social dynamics exposed to higher-order interactions.

\subsection{Contagion processes}\label{contg}

In order to take into account the group interactions of different sizes, Iacopini et al.~\cite{iacopini2019simplicial} formulated a higher-order simplicial model of social contagion. The model incorporates both pairwise and higher-order contacts, and thus combines the essences of both simple and complex contagion processes. Simplicial structure leads to a discontinuous transition to the endemic state and bistability emerges in which endemic and healthy states coexist. The former scenario has been demonstrated analytically along with numerical demonstrations on random Erd\H{o}s-R\'{e}nyi model and empirical higher-order networks. This model is then extended to the framework of temporal networks~\cite{chowdhary2021simplicial}, in which dyadic and higher-order interactions can be formed and destroyed temporally. Going through a microscopic Markov chain approach it has been shown that the same number of infectious seed may or may not generate endemic state, which actually depends on the temporal properties of the underlying network. The impact of degree heterogeneity on the simplicial contagion over time-varying higher-order networks is also investigated in the article thereafter. Social contagion dynamics is further investigated on hypergraphs in the Ref.~\cite{de2020social}. The authors particularly embodied the critical-mass dynamics into the previously framed model of the Ref.~\cite{iacopini2019simplicial}. Analytical and numerical results are presented to show the emergence of continuous and discontinuous transitions together with bistable regimes and hysteresis.

\par In addition to depicting that the standard network generating algorithms with tunable clustering characteristic can yield diverse higher-order structures so that dynamics can differ on the networks with same clustering and degree distribution profiles, the Ref.~\cite{ritchie2014higher} formulated a new metric for measuring order-four structures. The authors emphasize on how the higher-order structural differences (arising in networks possessing same clustering) have consequences on epidemic prevalence and epidemic threshold while dealing with Susceptible-Infected-Susceptible (SIS) and  Susceptible-Infected-Recovered (SIR) dynamical models. Landry et al.~\cite{landry2020effect} studied the dynamics of SIS model on hypergraphs by means of hyperdegree-based mean-field analysis on networks with higher-order interactions. Both degree-correlated and uncorrelated cases are analyzed while it has been unraveled that the abrupt first-order transitions can be suppressed through heterogeneous degree distribution of the dyadic interactions under certain assumptions on degree correlations. Besides inferring the conditions for bistability and hysteresis, the issues related to higher-order healing and the `hipster effect' have been further discussed in their article.

\begin{figure*}[ht]
	\centerline{\includegraphics[width = 16cm]{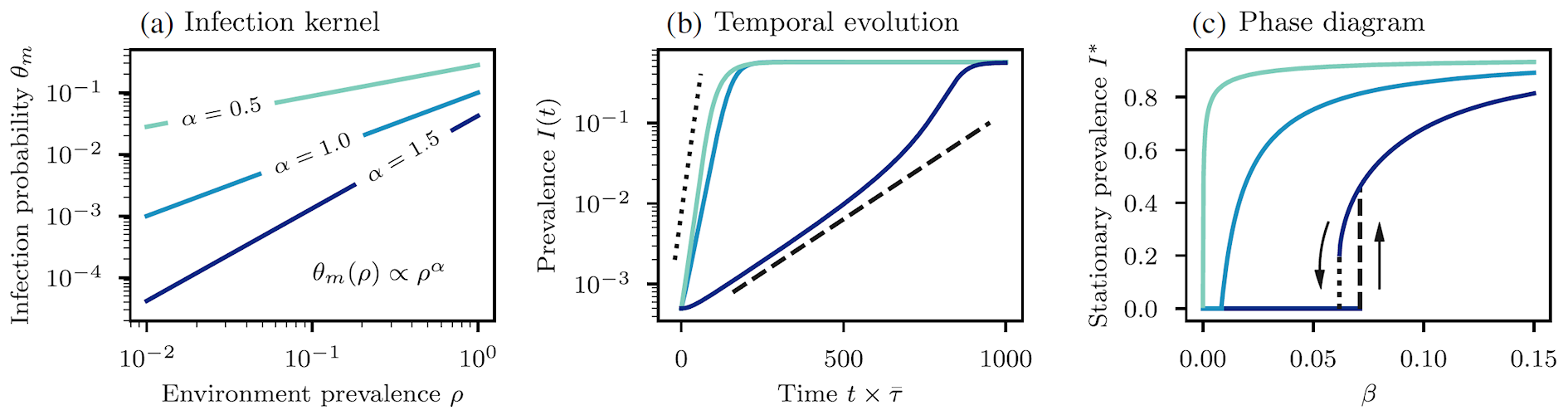}}
	\caption{(a) Infection kernel with $\beta=0.1$ where the infection probability ${\theta}_m$ has a power law scaling ${\theta}_m(\rho) \propto {\rho}^{\alpha}$. (b) Supralinear kernels $\nu > 1$ result in a superexponential growth of the prevalence $I(t)$. $\beta=0.0005,~0.025$ and $0.077$ for $\nu=0.5,~1.0$ and $1.5$ are chosen respectively, where $\bar{\tau}$ is the median exposure period. (c) Stationary prevalence $I^*$ with respect to $\beta$, where continuous phase transition appears for sublinear and linear kernels $\nu \le 1$ while a discontinuous phase transition arises with a bistable solution for supralinear kernels $\nu > 1$. Poisson distributions are opted for both the distribution $\bar{P}(m)$ of the size $m$ of the hyperlinks and the distribution $\tilde{P}(k)$ of hyperdegree $k$ of the individuals with $\langle m \rangle=10$ and $\langle k \rangle=5$ respectively, and $\mu=0.05$. Figure reproduced from the Ref.~\cite{st2021universal}.}
	\label{contag}
\end{figure*}

\par Lately, a higher-order model is developed addressing a number of issues that has been mostly neglected in epidemic modeling. Heterogeneity that arises in environments like offices and households etc., and the temporal heterogeneity in participation of the individuals in these environments are analyzed~\cite{st2021universal}. This heterogeneous exposure subject to minimal infective dose yields a universal nonlinear relation between the risk of infection and the infected contacts, challenging the prevalent assumption of linear relationship between these two. As a result, discontinuous transition to epidemic outbreak takes place and bistable regime emerges as well in which outbreak and healthy states coexist.  \\
Let us consider the interaction framework to be a hypernetwork in which the environments are defined by hyperlinks of $m$ individuals where each individual is incident to $k$ hyperlinks. A discrete-time process $(t=1,2,...)$ is then assumed where for each environment a participation time $\tau \in [1,~{\tau}_{\mbox{max}}]$ is chosen for each individual. Then if a susceptible individual participating in an $m$-sized environment for duration $\tau$ under the presence of a fraction $\rho$ of the other infected participants, it receives an infective dose $l \in [0,\infty)$ from the infected individuals, according to the distribution $f(l;\lambda)$, where $\lambda \equiv \langle l \rangle$. Similar to the threshold models, it is then considered that someone develops the disease when $l > K$ and ${\theta}_m(\rho)$ is the infection kernel that represents the he probability of getting the infection in an $m$-sized environment subject to a fraction $\rho$ of other infected participants. Also, $w \le {\tau}_{\mbox{max}}$ is the clearing window which represents the characteristic time for the immune system to be free of any dose $l<K$. Then, for heterogeneous exposure periods described by a Pareto distribution $P(\tau) \propto {\tau}^{-\alpha-1}$, $\alpha>0$ and the characteristic time to be infected as ${\tau}_c \equiv \frac{K}{\beta g(m)\rho}$, ($g(m)$ governs the number of contacts frequented by $m$ nodes and $\beta$ is a dose accumulation rate), ${\theta}_m(\rho)$ takes the form ${\theta}_m(\rho) \sim D_{\alpha} {{\tau}_c}^{-\alpha} \propto {\rho}^{\alpha}$~\cite{st2021universal}, with $D_{\alpha}$ being a constant. This is demonstrated in Fig.~\ref{contag}(a) for an exponential dose distribution $f(l;\lambda)=e^{-l/\lambda}$, $g(m)=1$ and $w \rightarrow \infty$ for different values of $\alpha$. \\
Further, the effects of nonlinear infection kernels are discussed while considering a SIS model with recovery rate being $\mu$. With a degree-based mean-field approximation, for the marginal probability ${\rho}_k(t)$ of an individual to be infected at time $t$ and $\tilde{P}(k)$ as the distribution of hyperdegree $k$ of the individuals, the global prevalence is $I(t)=\sum_{k}^{}{\rho}_k(t)\tilde{P}(k)$. The temporal evolution of this prevalence is portrayed in Fig.~\ref{contag}(b) that mirrors the impact of the nonlinear infection kernel. Specifically, superexponential growth is observed whenever $\nu > 1,~({\theta}_m(\rho) \propto {\rho}^{\nu})$ whereas the growth is regular exponential until saturation if $\nu \le 1$. Finally, Fig.~\ref{contag}(c) depicts the stationary prevalence $I^*$ as a function of $\beta$. The absorbing state $I^*=0$ remains unstable (cf. the dashed line in Fig.~\ref{contag}(c)) whenever $\beta> {\beta}_u$ (the invasion threshold). On the other hand, $I^*=0$ is globally stable (cf. the dotted line in Fig.~\ref{contag}(c)) for $\beta<{\beta}_s$ (the persistence threshold). It is conspicuous from the figure that the transition of $I^*$ can be either continuous (${\beta}_s={\beta}_u$) or discontinuous (${\beta}_s < {\beta}_u$) with a bistable solution.

\par A simplicial complex environment of interaction can again result in a discontinuous transition to the endemic state  ~\cite{li2021contagion}. In particular, here two different facets of contagion has been encountered, at the initial stage governed by the dyadic interactions whereas the later stage is controlled by the higher-order interactions. Theoretical analysis is provided in the homogeneous mixing limit along with rigorous computation in order to explain the associated bistable regime. By now, we all are aware of the fact that in case of contagions over standard network models built upon dyadic interactions, hubs play quite crucial roles. However, in higher-order networks, not only the individuals but also the groups play decisive roles. So, with the view of exploring the roles of set of groups on the hypergraph contagion dynamics, considering heterogeneity in both hyper-degree and hyperlink cardinality, in the Ref.~\cite{st2021influential} the authors construct a framework based upon approximate master equations analyzing contagion dynamics on top of random higher-order networks. Assuming the rate of infection as a nonlinear function of the number of infectious individuals in groups, it is shown how influential groups can govern the initial dynamics as well as the final stationary state of the contagion. A mathematical formulation has been provided with the view of analyzing the linear stability of general dynamical processes on arbitrary hypernetworks on the basis of weighted-graph projection of the hypernetwork~\cite{de2021phase}. In particular, the processes of social contagion and diffusion dynamics are dealt with. Apart from these, the study in the Ref.~\cite{li2022competing} investigated two competing SIS epidemic dynamics on a higher-order networked system composed of $1$- and $2$-simplex. Rigorous computations and the analysis of mean-field equations depicted a repertoire of dynamical features owing to the higher-order interaction. From the absolute dominance of the epidemics for weak triadic infection strength to the alternative dominance for higher triadic infection strength depending on the initial seed of infection are observed.

\subsection{Consensus formation}\label{cons}

Consensus dynamics on higher-order networked systems ($3$-body systems, mainly) is analyzed analytically and numerically~\cite{neuhauser2020multibody} in which it was disclosed that the dynamical consequences of multibody interactions can be effective only when interaction function is nonlinear. As a result of bringing in a nonlinear function herein, the emerged dynamics causes shifts off the state of average system, depending on the underlying network and the initial configuration. Consensus dynamics in higher-order networks of any order is further studied in the Ref.~\cite{sahasrabuddhe2021modelling} while contemplating with a number of social processes like homophily and peer-pressure for modeling the interactions. Apart from the hypergraph models like block hypergraphs, analysis has been performed on real-world networks as well. In~\cite{hickok2021bounded}, the authors formulated a hypergraph bounded confidence model and showed the appearance of a scenario named as `opinion jumping' in which individuals' opinion can jump from one cluster of opinions to another, which one doesn't observe in dyadic connectivity. Moreover, echo chambers are witnessed to emerge on hypergraphs with community structure. Large hyperlinks are found to be playing decisive roles for the consensus more than the small hyperlinks. Besides the computational demonstrations, the scenarios are treated mathematically. Consensus dynamics over higher-order networked systems can be investigated through the concept of \textit{generalized Hodge Laplacians} for the instances in which the weights for lower- and higher-order interactions between simplices are different~\cite{ziegler2021balanced}. Using the Hodge Decomposition, convergence can be analyzed and thereafter with the techniques of algebraic topology the role of simplicial complex homology can be studied. In fact, lower- and higher-order interactions can be balanced to optimize consensus dynamics.

\begin{figure*}[ht]
	\centerline{\includegraphics[width = 16cm]{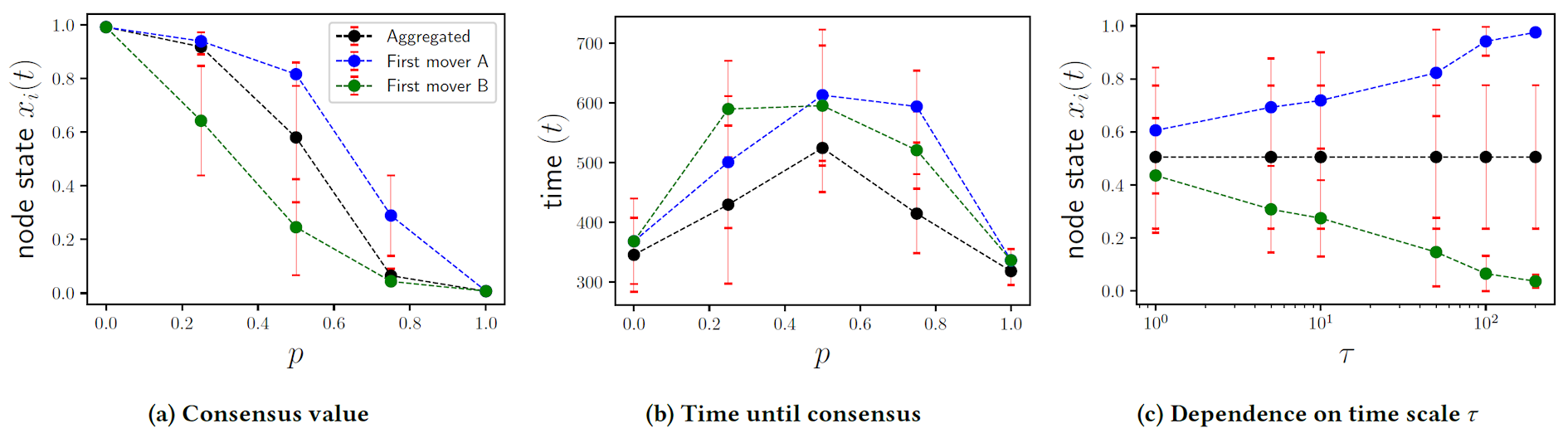}}
	\caption{(a) Consensus value with respect to the parameter $p$ corresponding to the scenarios of first-mover A, first-mover B and the aggregated dynamics, in which the error bars reflect one standard deviation. (b) Time taken to reach consensus as a function of the fraction $p$ that shows faster rate of convergence for asymmetric configuration. (c) Time-scale $\tau$ dependence of the consensus value for $p=0.5$. Decreasing $\tau$ helps in making the convergence to the aggregated dynamics. Figure reproduced from the Ref.~\cite{neuhauser2021consensus}.}
	\label{consen}
\end{figure*}

\par In the above, we have already discussed how temporal higher-order interaction patterns modulate the discrete dynamics of social contagion~\cite{chowdhary2021simplicial}. Let us now elaborate how temporality in network connectivity affects the continuous dynamics of consensus process developing in higher-order networks~\cite{neuhauser2021consensus}. \\
The nodal dynamics is described by the following set of equations:
\begin{equation}
	\begin{array}{lcl}\label{eq1}

		\dot x_i=\displaystyle\sum_{j,k}^{}A_{ijk}\mbox{exp}(l|x_k-x_j|)\Big[(x_j-x_i)+(x_k-x_i)\Big],\\
		~~~~~~~~~~~~~~~~~~~~~~~~~~~~~~~~~~~~~~i=1,2,\cdots ,N.

	\end{array}
\end{equation}
where $A_{ijk} \in {\mathbb{R}}^{N \times N \times N}$ is the adjacency tensor representing the interaction structure of the $3$-hypernetwork and the term $\mbox{exp}(l|x_k-x_j|)$ is the scaling function that regulates the impacts of $j$-th and $k$-th nodes on the $i$-th node. Then the temporal network model of $3$-regular hypergraphs is constructed by defining a sequence of adjacency tensors $A^{[1]},~A^{[2]},...$ representing the network structures at different times with $\tau$ being the length of the time-periods between any two successive adjacency tensors.
\par Let us then assume a network set-up with two (individually globally connected) clusters (say, cluster-`A' and `B') of same size ($N=10$ nodes) in which both intra-cluster and inter-cluster hyperlink connections exist with the nodes in cluster A (B) having the initial state $x_A(0)=1$ ($x_B(0)=0$). Further, the clusters are connected via $20$ randomly placed hyper $3$-links in such a way that $p \in [0,~1]$-fraction of $3$-links are oriented towards the cluster A (i.e., lesser number of hyperlink nodes are part of the cluster A than of the cluster B) and the rest of the $3$-links are oriented towards B. Then three different schemes are studied, namely the first-mover A, first-mover B and the aggregated scenario. To be precise, in case of first-mover A (first-mover B), firstly for certain time all the A-majority (B-majority) subgroups interact, then all the B-majority (A-majority) subgroups and then the entire hypernetwork interacts. For the aggregated scenario, the hypernetwork remains static and all the interactions take place concurrently. \\
Figure~\ref{consen}(a) portrays the value of consensus in terms of the node state $x_i(t)$ as a function of the fraction $p$, averaged over $10$ simulations. The consensus evolving in the hypernetwork tends towards the initial opinion in cluster A or B for the aggregated scenario. Whenever $p=0$ ($p=1$), the connecting $3$-links are oriented towards cluster B (A) making the initial opinion of cluster A (B) prevail. This outcome turns out to be qualitatively same in the first mover cases. As far as the convergence speed is concerned, convergence is effectively faster for asymmetric initial opinion and when orientation of the $3$-links and the first-mover group line up (cf. Fig.~\ref{consen}(b)). Further, Fig.~\ref{consen}(c) demonstrates the consensus dynamics with respect to the time-scale $\tau$ for $p=0.5$. It is clear from the figure that the system is more prone to converge to the aggregated dynamics whenever $\tau$ is small.

\par In the Ref.~\cite{horstmeyer2020adaptive}, the authors propose and study an adaptive voter model under higher-order interactions, specifically on a simplicial complex, that incorporates the influence of the important social factor of peer-pressure. The rewiring rule of linking to agreeing nodes is adopted while focusing mainly on the $2$-simplex framework. Peer-pressure speeds up the transitions to both single-opinion and two-opinion states. Also, this higher-order model can exhibit multiple time-scales in which the $2$-simplices vanish before the exhaustion of the active links.  In a recent work~\cite{deville2021consensus}, a generalized dynamical model on simplicial complex of several consensus and synchronization processes is proposed and analyzed. Many behaviors are detected here for consensus dynamics that occur for dyadic interactions and also the emergence of multistability in the steady states due to this model is put forward.

\subsection{Evolutionary game dynamics}\label{egd}

\begin{figure*}[ht]
	\centerline{\includegraphics[width = 12cm]{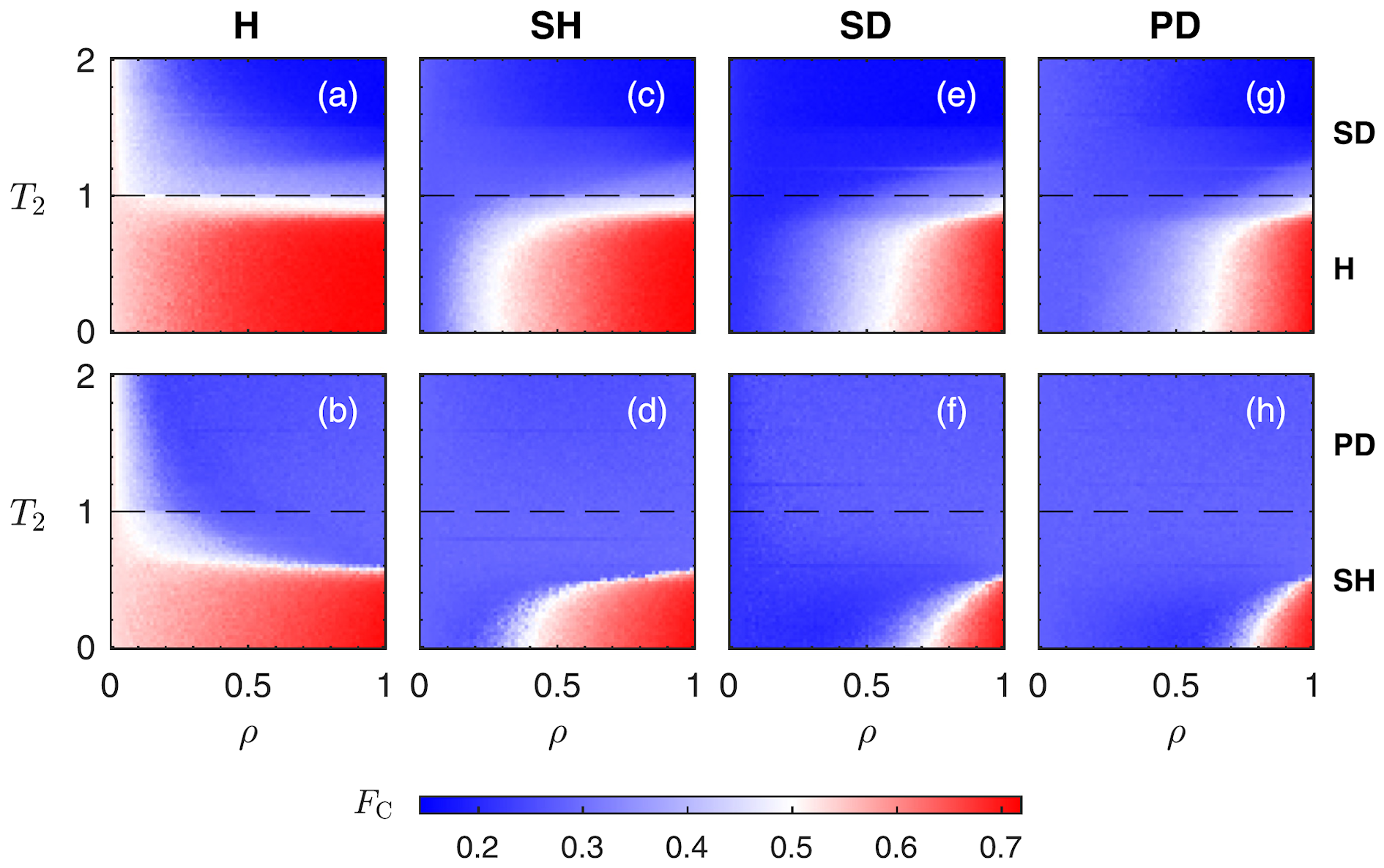}}
	\caption{ $\rho-T_2$ phase diagrams depicting the frequency $F_C$ of cooperators, for fixed values of $S_2=0.5$ for the first row and $S_2=-0.5$ for the second row, with (a,b) $T_1=T_3=0.8,~S_1=S_3=0.2$ that correspond to the Harmony game, (c,d) $T_1=T_3=0.8,~S_1=S_3=-0.2$ that represent the Stag Hunt game, (e,f) $T_1=T_3=1.2,~S_1=S_3=0.2$ that define the Snowdrift game, and (g,h) $T_1=T_3=1.2,~S_1=S_3=-0.2$ associated to the Prisoners Dilemma game. In the first row, Game2 is SD for $T_2 \geq 1$ and H dilemma when $T_2 \leq 1$, while in the second row, Game2 is PD if $T_2 \geq 1$ and a SH dilemma for $T_2 \leq 1$. Figure reproduced from the Ref.~\cite{guo2021evolutionary}.}
	\label{game}
\end{figure*}

Cooperation ~\cite{nowak2006five,perc2017statistical,perc2010coevolutionary,szolnoki2014cyclic,hilbe2014cooperation,perc2013evolutionary,perc2008social} is the process in which individuals function together in groups for mutual benefits, which is observed in diverse real systems including microorganisms and human society. Significant attempts have been made earlier in order to explore the evolutionary game dynamics in populations subject to group interactions (cf.~\cite{perc2013evolutionary} and references therein). In evolutionary game theory, the higher-order interactions differ from pairwise interactions in the derivation of payoffs. If one's
payoff in a higher-order interaction, to some degree, is equivalent to the sum of payoffs in interactions with each individual opponent, both higher-order and pairwise interactions	essentially are the same. Otherwise, if one's payoff in his neighborhood is nonlinear to the sum of payoffs in all pairwise interactions, higher-order interactions lead to different dynamical	processes. An approach to capture the higher-order interaction is general multiplayer game, where one player's payoff is a function of his and all neighbors' strategies (the Refs.~\cite{li2014cooperation,pena2016evolutionary,li2016evolutionary,zhou2018simple,su2019evolutionary,su2018understanding,su2019spatial} furnish nice strict analytical results on multiplayer games). When the payoff function is nonlinear to the number of cooperative neighbors, it presents the higher-order effects.

\par In particular, the review by Perc et al.~\cite{perc2013evolutionary} clarifies how larger group sizes can help in preserving cooperation in networks formed upon dyadic interactions which are often insufficient to explain all the essence of group interactions. Keeping this in mind, Burgio et al.~\cite{burgio2020evolution} came up with their work on diverse hypernetworks in pursuance of having clearer perception of the development of cooperation in networked groups while examining the evolution of cooperation in public goods game (PGG), and demonstrated that group interactions can, indeed, enhance cooperation. The method adopted to generate the hypernetworks preserve the dyadic projection and the authors, in particular, deal with hypernetworks formed from the Holme-Kim and the Dorogovtsev-Mendes models. Besides mean-field approximation for homogeneous interactions, invasion analysis is presented for heterogeneous structures explaining how increasing order of connections can cause higher reciprocity. The developed reciprocity specifically owes to the adopted mechanism that replaces some first-order $3$-cliques with second-order triangles. The article also discusses how cooperative and non-cooperative states can coexist subject to the modality of interaction structures.
\par Evolutionary dynamics of public goods game is also investigated in social networks built upon higher-order interactions~\cite{alvarez2021evolutionary}. The study reveals that this game on uniform hypernetworks in which there is no hyperdegree-hyperdegree correlations is consistent with the replicator dynamics in the well-mixed regime. The article further incorporates heterogeneity in both order and hyperdegree, and demonstrates how these characteristic features affect the evolutionary game dynamics. This higher-order network framework being capable of appropriately describing the group structures, the study actually has been able to depict how synergy factors depending on the group size result in critical scaling in the defection to cooperation transition. Hierarchical hypernetworks are observed to impede cooperation in a structured population. The network set-up is further employed to collaboration datasets as well. Higher-order group ($3$-player with $2$-player) interactions along with adaption are taken into account in order to propose the adaptive simplicial Snow-drift game~\cite{schlager2021stability}. Adaptation in the network topology and the state of the system is assumed which explores for both mathematical and numerical treatments that even under the higher-order structural framework, the stability of the equilibrium points remains unaltered. An evolutionary model of group choice dilemmas is proposed and analyzed on hypernetworks where the decisions between a safe alternative and a risky one are taken in different sized groups, which is capable of explaining how opinion diffuses following an imitation process~\cite{civilini2021evolutionary}. Further, an organized study for a different form of strategic interaction of signaling games in populations subject to higher-order structures, namely the dynamical evolution of honesty in the sender-receiver game is presented very recently~\cite{kumar2021evolution}. Unlike the instance of the sole presence of dyadic interactions, honesty has been witnessed to be existent even under the temptation to lie. Also, moral strategy persists even if lies favor the receiver at a cost to the sender. The evolutionary dynamics is investigated in populations based upon the assumption of well-mixed setting, in hyper-ring as well as in real-world hypernetworks.

\par A different approach is adopted very recently to model evolutionary game dynamics for higher-order interactions among individuals, where apart from the strategies of a focal player and one of the neighbors, strategies of other neighbor(s) coming out of an indirect interactions also influence the game dynamics~\cite{guo2021evolutionary}. Diverse social dilemmas with different Nash equilibria being played over $1$- and $2$-simplices are investigated demonstrating that such a simplicial framework results in the appearance of the non-dominant strategies and its coexistence with the dominant strategies. Further, transition from dominant defection state to the state of cooperation with respect to the higher-order structure is established. \\
The two-strategy (cooperation (C) and defection (D)) two-player game configuration can be described by the following payoff matrix:
\begin{equation}
	\begin{array}{lcl}\label{eq2}	
			& C & D \\
			C & R & S \\
		
			D & T & P
	\end{array}
\end{equation}
Each player receives a payoff $R=1$ (Reward) under mutual cooperation and $P=0$ (Punishment) for mutual defection on the agreement of the strategies. Instead, if the players' strategies disagree, the cooperator receives a payoff $S \in [-1,1]$ (Sucker) whereas the defector receives $T \in [0,2]$ (Temptation). Concerning the network formulation, initially starting with a fully connected sub-network of $n_0(=5)$ nodes, in the next time-step $m(=1)$ new nodes are added. These new nodes are linked to the endpoints of randomly chosen $m$ links, and thus $m$ new triangles are created in the sub-network. Reiterating this step of addition of nodes, the final network of $N$ nodes is constructed. The network thus formed exhibits power-law degree distribution with exponential generalized-degree distribution~\cite{kovalenko2021growing}. A fraction $\rho \in [0,1]$ of random triangles in the network are chosen to characterize actual three-body ($2$-simplex) interactions whereas the remaining fraction ($1-\rho$) of triangles represents three two-body ($1$-simplex) interactions. A strategy matrix $\tilde{S}=\{s_{ij}\}$ is also defined that takes different values based on whether $i$-th and $j$-th nodes cooperate, defect or do not interact. The accumulated payoff $\Pi_i$ of the $i$-th node is then calculated as $\Pi_i=(1/k_i)\sum_{j \in N_i}^{}\Pi_{i,(ij)}$, where $N_i,~k_i$ and $\Pi_{i,(ij)}$ are respectively the neighbourhood, degree and the total payoff obtained along the link $(i,j)$ of the $i$-th node. Further, $\Pi_{i,(ij)}=(1/k_{ij})\sum_{\tau \in \Delta}^{}\Pi_{i,(ij),\tau}$, where the set $\Delta$ comprises of the $k_{ij}$ triangles constituted by the link $(i,j)$. Now, if $\tau$ is simply a sum of three $1$-simplices, then $\Pi_{i,(ij),\tau}$ is obtained from Game1, the payoff values of which are $S=S_1,~T=T_1$ with $R=1,~P=0$. On the other hand, if $\tau$ characterizes a $2$-simplex, then assuming the other node to be the $k$-th node that completes this simplex, $\Pi_{i,(ij),\tau}$ will be calculated from Game2 ($S=S_2,~T=T_2$ with $R=1,~P=0$) if $s_{ki}=s_{kj}$. Similarly, the payoff will be obtained from Game3 ($S=S_3,~T=T_3$ with $R=1,~P=0$) if $s_{ki} \neq s_{kj}$. This way each $i$-th node obtains its payoff $\Pi_i$ and subsequently updates (synchronously with others) its strategy with probability $\mbox{Pr}=1/(1+e^{[(\Pi_i-\Pi_{\tilde{j}})/K]})$, in which $\Pi_{\tilde{j}}$ is the accumulated payoff of the $\tilde{j}$-th node.

\par Frequency of cooperation in the $\rho-T_2$ parameter plane is shown in Fig.~\ref{game} while considering Game1 and Game3 to be the same (i.e., with the same $S$ and $T$ values). Game1,3 are identified respectively by the Harmony (H), Stag Hunt (SH), Snowdrift (SD) and Prisoner's Dilemma (PD) games from the left to the right columns of the figure. Moreover, for the upper (lower) row $S_2=0.5$ ($S_2=-0.5$) is assumed so that the Game2 represents H (SH) dilemma whenever $T_2 \leq 1$ and SD (PD) game for $T_2 \geq 1$.  For the left-most Harmony dilemma (cf. Figs.~\ref{game}(a) and~\ref{game}(b)), mutual cooperation being the Nash equilibrium, for small values of $\rho$ cooperation is favored irrespective of the specifics of Game2. With increasing $\rho$, the number of three-body interactions increases and the Game2 dynamics starts to matter which can be any of the considered four game dynamics. For $T_2 \leq 1$, Game2 is either H or SH that supports cooperation and hence $\rho$ values do not matter much. However, whenever $T_2 \geq 1$, Game2 is SD or PD and hence the fraction of cooperators decreases. Interesting results start appearing when Game1 and Game3 correspond to SH, SD and PD (cf. Figs.~\ref{game}(c-h)). In absence of simplicial interactions (i.e., $\rho=0$), defection is the dominant strategy but a transition to cooperation takes place with increasing $\rho$ (implying increasing higher-order interactions) whenever $T_2 <1$ (i.e., Game2 is either H or SH). More than the transition scenario for Game2 playing the Harmony game (for which CC is the Nash equilibria), the transition to cooperation (even when Game1 and Game3 represents PD) for the instance of Game2 playing the SH game is noteworthy as CC and DD are the two pure Nash equilibria for the SH game.

\section{Random walk and diffusion}\label{rand}

With an aim to explore the dynamics of random walks on networks beyond pairwise interactions, a family of random walks on top of simplicial complexes is defined by Markov chain~\cite{mukherjee2016random}. A relation between the chain's stationary distribution and the harmonics of the Hodge Laplacian is further established. In this context, from the higher-order homology groups and the role of orientation of the simplexes to the concept of neighbors of the higher-order simplexes are discussed in detail. Also in the Ref. ~\cite{parzanchevski2017simplicial}, the authors have gone through the concept of random walks on simplicial complexes. Diffusive processes in the form of a family of random walks on heterogeneous higher-order networks (hypernetworks) is brought forward while giving an analytical treatment with a general proposition for the stationary distribution of the walk~\cite{carletti2020random}. A comparison analysis of this distribution to that corresponding to the traditional random walk over the associated projected network is also provided. Both model and real-world hypernetworks are treated in order to explore the proposed random walk dynamics. Specifically, from the applications in node ranking and centrality measure to classification tasks are explained. The process of diffusion on simplicial complexes is studied in the Ref.~\cite{schaub2020random} that proposes a normalized Hodge Laplacian matrix and demonstrates how it is associated to random walk dynamics on simplicial complexes, specifically on edges. The approach is further utilized in developing embeddings of edge flows and trajectory data and also the generalization of personalized PageRank for edges.

\begin{figure*}[ht]
	\centerline{\includegraphics[width = 11.4cm]{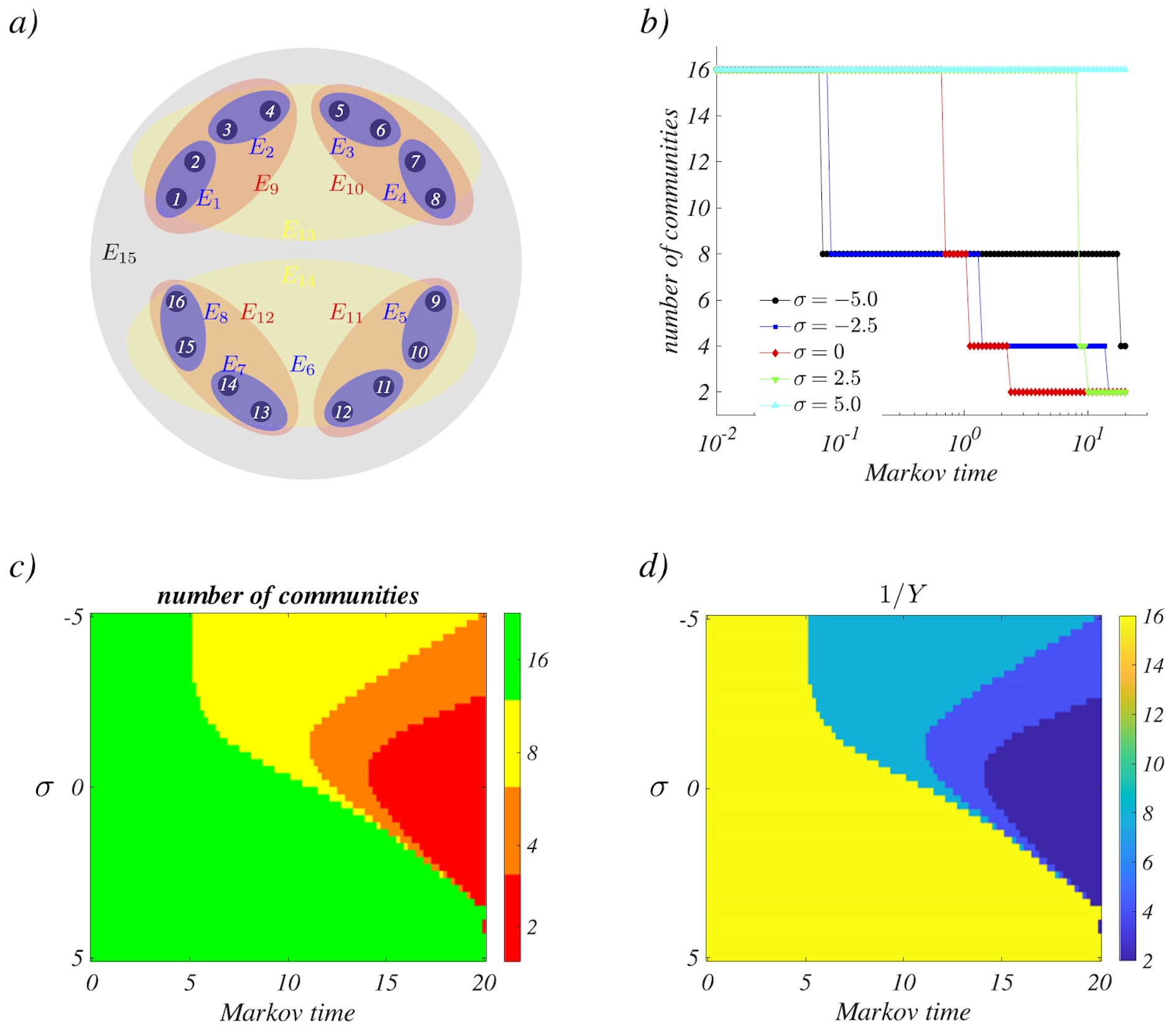}}
	\caption{ (a) The hypernetwork composed of $16$ nodes and $15$ hyperlinks, specifically there exists $8$ hyperlinks of size $2$ (blue), $4$ hyperlinks of size $4$ (red), $2$ hyperlinks of size $4$ (yellow) and $1$ large hyperlink of size $16$ (grey). (b) The number of communities as a function of the Markov time for different values of the size bias parameter $\sigma$. (c) Number of communities for simultaneous variations of Markov time and $\sigma$. (d) $1/Y$ with respect to Markov time and $\sigma$. Figure reproduced from the Ref.~\cite{carletti2021random}.}
	\label{random}
\end{figure*}

\par A class of random walks on hypernetworks is defined in such a way that the random walk process shows propensity towards hyperlinks of high or low size based upon the variation of a single size bias parameter~\cite{carletti2021random}. The resulting dynamics is, in fact, capable of describing diverse hypernetwork projection on networks for different values of this bias parameter. These projections can further vary depending on this parameter and this dissimilation is examined via its effect on community structure while developing the formulation of Markov stability on hypernetworks. Let us assume a hypernetwork $H(V,E)$ with $V=\{V_1, V_2,..., V_N\}$ and $E=\{E_1, E_2,..., E_M\}$ being the sets of $N$ nodes and $M$ hyperlinks respectively. The incidence matrix associated to the hypernetwork is the following:
\[
{e}_{i\alpha}=
\begin{cases}
	1, & \text{if}~V_i \in E_{\alpha}, \\
	0,              & \text{otherwise}.
\end{cases}
\]\\
The $M \times M$ hyperlink matrix is defined as $\bf{B}={\bf{e}}^t{\bf{e}}$ in which ${\bf{e}}^t$ is the transpose of ${\bf{e}}$ and the elements $B_{\alpha\beta}$ accounts for the number of nodes in $E_{\alpha} \cap E_{\beta}$. The agents are then placed on the nodes that hop at discrete times, and the weighted adjacency matrix is described as ${K_{ij}}^{[\sigma]}=\sum_{\alpha}^{}{(B_{\alpha\alpha}-1)}^{\sigma}{e}_{i\alpha}{e}_{j\alpha},~\sigma \in \mathbb{R},~\forall i\neq j$ and ${K_{ii}}^{[\sigma]}=0$, from which the transition probabilities are computed as ${T_{ij}}^{[\sigma]}={K_{ij}}^{[\sigma]}/(\sum_{m \neq i}^{}{K_{im}}^{[\sigma]}),~\forall i\neq j$ and ${T_{ii}}^{[\sigma]}=0$. This implies how the hyperlinks of large (small) size governs the random walk dynamics for large (negative) values of the size bias parameter $\sigma$. A continuous random walk on top of the hypernetwork is then delineated as

\begin{equation}
	\begin{array}{lcl}\label{eq3}
		
		\dot p_i=\displaystyle\sum_{j}^{}p_{j}{T_{ji}}^{[\sigma]}-\displaystyle\sum_{j}^{}p_{i}{T_{ij}}^{[\sigma]},~~~~~~~~i=1,2,\cdots ,N,
		
	\end{array}
\end{equation}
in which $p_i~(p_i(t))$ is the probability of the agent being on the $i$-th node at time $t$. \\
A generalization of the formulation of Markov stability~\cite{delvenne2010stability} is further adopted in order to find the communities in the hypernetwork, by assuming a partition of the nodes into $c$ non-overlapping communities, captured by the indicator matrix ${\bf C}_{N \times c}$ where $C_{ij}$ takes up the value $1$ when the $i$-th node belongs to the $j$-th community, and $0$ otherwise. The Markov stability $r(t;{\bf C})$ then measures the goodness of ${\bf C}$ as a function of the time horizon of the random walk (please see the Ref.~\cite{carletti2021random} for detailed definitions of Markov Stability).

\par We then consider a typical hierarchical hypernetwork model that comprises of $16$ nodes and $15$ hyperlinks (cf. Fig.~\ref{random}(a)), the projection of which is a complete network with $16$ nodes. As can be seen, there exists $8$ hyperlinks each of size $2$, $4$ hyperlinks containing $4$ nodes, $2$ hyperlinks with $8$ nodes and lastly the hyperlink containing all the $16$ nodes. In order to find the communities, Markov stability is optimized with respect to the Markov time $t$ for different values of $\sigma$ in Fig.~\ref{random}(b). The plots demonstrate the hierarchical structure efficiently while determining all the communities of decreasing sizes as the Markov time increases. Then having a look at the entries of ${\bf K}^{[\sigma]}$ one can calculate $\lim_{\sigma \to +\infty} {T_{ij}}^{[\sigma]}=1/15,~\forall i,j \in \{1,2,...,16\}$, with $\lim_{\sigma \to -\infty} {T_{12}}^{[\sigma]}=1$ and $\lim_{\sigma \to -\infty} {T_{1j}}^{[\sigma]}=0$ for all other $j$. The other values follow from the symmetry in the structure of the hypernetwork. Figure~\ref{random}(c) depicts the number of communities as a function of Markov time and $\sigma$. The algorithm yields the partition of $16$ communities whenever $\sigma$ is positive and large, and the two communities of size $8$ for smaller positive $\sigma$. The intermediate communities are realized mainly for the negative values of $\sigma$, and the algorithm finally identifies the communities of size $2$. Furthermore, the Simpson diversity index $Y$ is computed in order to get the size of the communities, where  $Y=\sum_{i=1}^{Q}{S_i}^2/N^2$, in which $S_i$ is the number of nodes present in the $i$-th group. $Y$ varies from $1$ (when all the nodes are in a single group) to $1/N$ (if there exists $Q=N$ groups, each comprising of a single node), whereas $Y \sim 1/Q$ whenever the nodes are uniformly distributed among the $Q$ groups. In Fig.~\ref{random}(d), the value of $1/Y$ is presented for simultaneous variations in Markov time and $\sigma$. It is discernible that $1/Y=2$ is associated to the $Q=2$ communities of size $8$ as here $Y=2 \times {(8/16)}^2=1/2$. Similarly, $1/Y=4$ corresponds to the $Q=4$ communities of size $4$, and $1/Y=8$ represents $Q=8$ communities of size $2$.

\par  Spectral properties of higher-order Laplacian associated to the simplicial complex model known as `network geometry with flavor' (as mentioned above) are studied in the Ref.~\cite{torres2020simplicial}. It is demonstrated that these higher-order up- and down-Laplacians can have a finite spectral dimension that depends on the order of the Laplacian. Moreover, this higher-order structure affects the diffusion dynamics taking place on this, with the spectral dimensions having meaningful influence on the return-time probability of the concerned diffusion process. Furthermore, relation between the geometry of a network and diffusion dynamics is unraveled~\cite{millan2021local} based on the investigation of two family of models, namely NGF and `Short-Range Triadic Closure (STC)'. Thus far, many generalizations of different random walk models for higher-order interactions have been put forward, as discussed above. In order to explore which combination of model and network representation is best for resolving different research issues associated with diverse hypernetwork data, the authors of the Ref.~\cite{eriksson2021choosing} derive unipartite, bipartite and multilayer network representations of hypernetwork flows with identical node-visit rates for the same random-walk model. Information-theoretic and flow-based community detection algorithm Infomap is utilized to investigate how different hypernetwork models and network representations alter the number, size, overlap of the detected communities.

\section{Summary and Future Prospects}\label{summ}

The variants of interactions in networked systems essentially regulate the dynamical processes taking place on them. It has been demonstrated in many ways that from synchronization to spreading dynamics, the complex interaction structure strongly decides the destiny of the concerned complex systems. However, the existing literature predominantly has dealt with pairwise networked systems, even though the underlying processes are better represented on top of higher-order structures. Only in the recent times, structural and dynamical properties of higher-order networks have become a rapidly developing research field owing to its potential efficacy in describing numerous complex instances from social processes to neuroscience. In this review article, we have furnished a review of recent research endeavours that study various dynamical processes on networks beyond dyadic interactions. We come up with our investigation with a motive to clarify how diverse can the impact on different phenomena be while higher-order connections are taken into account. The fundamental concepts of higher-order networks are briefly discussed in Section II. In Section III, we started with explaining how the phenomenon of synchronization gets affected due to the presence of higher-order connections in the system. Social processes staring from contagion dynamics, consensus formation to evolution of cooperation are examined in Sections IV A, IV B and IV C, respectively. The influence of higher-order interactions on random walk and diffusion dynamics are studied in Section V.

Even though a number of significant developments have been made in view of analyzing the role of higher-order interactions on dynamical processes, we would still like to bring forward some of the noteworthy routes of further research. For instance, there are enough scopes to contribute to the understanding of the temporal higher-order networks. From its structural intricacies to the analysis of different dynamics on time-varying higher-order structures is highly worth of attention. The same applies to the interdependent network frameworks, specifically the multilayer/multiplex structures along with higher-order interactions, inspection of which should be envisaged as a promising research direction. Although there exists important attempts concerning synchronization in networks beyond pairwise connectivity, the detailed analysis of cluster synchrony is missing. The specific aspect of chimera state is mostly untouched so far whereas these patterns have high resemblance to several neuronal developments~\cite{majhi2019chimera}. So, the study of chimera states in simplicial networks would be an excellent candidate for future research. Also, the study of collective behaviors of swarmalator systems with higher-order interactions could be quite interesting. Moreover, the dynamical scenarios arising from the increased complexity due to adaptivity~\cite{gross2008adaptive} in the higher-order systems require much more attentive study.

\section*{\normalsize{Acknowledgments}}
S.M. acknowledges support from the Planning and Budgeting Committee of Israel's Council for Higher Education. M.P. acknowledges funding from the Slovenian Research Agency (Grant nos. P1-0403 and J1-2457).

\providecommand{\noopsort}[1]{}\providecommand{\singleletter}[1]{#1}%


\begin{thebibliography}{128}
\expandafter\ifx\csname natexlab\endcsname\relax\def\natexlab#1{#1}\fi
\expandafter\ifx\csname bibnamefont\endcsname\relax
  \def\bibnamefont#1{#1}\fi
\expandafter\ifx\csname bibfnamefont\endcsname\relax
  \def\bibfnamefont#1{#1}\fi
\expandafter\ifx\csname citenamefont\endcsname\relax
  \def\citenamefont#1{#1}\fi
\expandafter\ifx\csname url\endcsname\relax
  \def\url#1{\texttt{#1}}\fi
\expandafter\ifx\csname urlprefix\endcsname\relax\def\urlprefix{URL }\fi
\providecommand{\bibinfo}[2]{#2}
\providecommand{\eprint}[2][]{\url{#2}}

\bibitem[{\citenamefont{Albert and Barab{\'a}si}(2002)}]{albert2002statistical}
\bibinfo{author}{\bibfnamefont{R.}~\bibnamefont{Albert}} \bibnamefont{and}
  \bibinfo{author}{\bibfnamefont{A.-L.} \bibnamefont{Barab{\'a}si}},
  \bibinfo{journal}{Reviews of Modern Physics} \textbf{\bibinfo{volume}{74}},
  \bibinfo{pages}{47} (\bibinfo{year}{2002}).

\bibitem[{\citenamefont{Newman}(2003)}]{newman2003structure}
\bibinfo{author}{\bibfnamefont{M.~E.} \bibnamefont{Newman}},
  \bibinfo{journal}{SIAM Review} \textbf{\bibinfo{volume}{45}},
  \bibinfo{pages}{167} (\bibinfo{year}{2003}).

\bibitem[{\citenamefont{Boccaletti et~al.}(2006)\citenamefont{Boccaletti,
  Latora, Moreno, Chavez, and Hwang}}]{boccaletti2006complex}
\bibinfo{author}{\bibfnamefont{S.}~\bibnamefont{Boccaletti}},
  \bibinfo{author}{\bibfnamefont{V.}~\bibnamefont{Latora}},
  \bibinfo{author}{\bibfnamefont{Y.}~\bibnamefont{Moreno}},
  \bibinfo{author}{\bibfnamefont{M.}~\bibnamefont{Chavez}}, \bibnamefont{and}
  \bibinfo{author}{\bibfnamefont{D.-U.} \bibnamefont{Hwang}},
  \bibinfo{journal}{Physics Reports} \textbf{\bibinfo{volume}{424}},
  \bibinfo{pages}{175} (\bibinfo{year}{2006}).

\bibitem[{\citenamefont{Barrat et~al.}(2008)\citenamefont{Barrat, Barthelemy,
  and Vespignani}}]{barrat2008dynamical}
\bibinfo{author}{\bibfnamefont{A.}~\bibnamefont{Barrat}},
  \bibinfo{author}{\bibfnamefont{M.}~\bibnamefont{Barthelemy}},
  \bibnamefont{and}
  \bibinfo{author}{\bibfnamefont{A.}~\bibnamefont{Vespignani}},
  \emph{\bibinfo{title}{Dynamical processes on complex networks}}
  (\bibinfo{publisher}{Cambridge university press}, \bibinfo{year}{2008}).

\bibitem[{\citenamefont{Boccaletti et~al.}(2014)\citenamefont{Boccaletti,
  Bianconi, Criado, Del~Genio, G{\'o}mez-Gardenes, Romance, Sendina-Nadal,
  Wang, and Zanin}}]{boccaletti2014structure}
\bibinfo{author}{\bibfnamefont{S.}~\bibnamefont{Boccaletti}},
  \bibinfo{author}{\bibfnamefont{G.}~\bibnamefont{Bianconi}},
  \bibinfo{author}{\bibfnamefont{R.}~\bibnamefont{Criado}},
  \bibinfo{author}{\bibfnamefont{C.~I.} \bibnamefont{Del~Genio}},
  \bibinfo{author}{\bibfnamefont{J.}~\bibnamefont{G{\'o}mez-Gardenes}},
  \bibinfo{author}{\bibfnamefont{M.}~\bibnamefont{Romance}},
  \bibinfo{author}{\bibfnamefont{I.}~\bibnamefont{Sendina-Nadal}},
  \bibinfo{author}{\bibfnamefont{Z.}~\bibnamefont{Wang}}, \bibnamefont{and}
  \bibinfo{author}{\bibfnamefont{M.}~\bibnamefont{Zanin}},
  \bibinfo{journal}{Physics Reports} \textbf{\bibinfo{volume}{544}},
  \bibinfo{pages}{1} (\bibinfo{year}{2014}).

\bibitem[{\citenamefont{Holme and Saram{\"a}ki}(2012)}]{holme2012temporal}
\bibinfo{author}{\bibfnamefont{P.}~\bibnamefont{Holme}} \bibnamefont{and}
  \bibinfo{author}{\bibfnamefont{J.}~\bibnamefont{Saram{\"a}ki}},
  \bibinfo{journal}{Physics Reports} \textbf{\bibinfo{volume}{519}},
  \bibinfo{pages}{97} (\bibinfo{year}{2012}).

\bibitem[{\citenamefont{Ghosh et~al.}(2022)\citenamefont{Ghosh, Frasca, Rizzo,
  Majhi, Rakshit, Alfaro-Bittner, and Boccaletti}}]{ghosh2022synchronized}
\bibinfo{author}{\bibfnamefont{D.}~\bibnamefont{Ghosh}},
  \bibinfo{author}{\bibfnamefont{M.}~\bibnamefont{Frasca}},
  \bibinfo{author}{\bibfnamefont{A.}~\bibnamefont{Rizzo}},
  \bibinfo{author}{\bibfnamefont{S.}~\bibnamefont{Majhi}},
  \bibinfo{author}{\bibfnamefont{S.}~\bibnamefont{Rakshit}},
  \bibinfo{author}{\bibfnamefont{K.}~\bibnamefont{Alfaro-Bittner}},
  \bibnamefont{and}
  \bibinfo{author}{\bibfnamefont{S.}~\bibnamefont{Boccaletti}},
  \bibinfo{journal}{Physics Reports} \textbf{\bibinfo{volume}{949}},
  \bibinfo{pages}{1} (\bibinfo{year}{2022}).

\bibitem[{\citenamefont{Battiston et~al.}(2020)\citenamefont{Battiston,
  Cencetti, Iacopini, Latora, Lucas, Patania, Young, and
  Petri}}]{battiston2020networks}
\bibinfo{author}{\bibfnamefont{F.}~\bibnamefont{Battiston}},
  \bibinfo{author}{\bibfnamefont{G.}~\bibnamefont{Cencetti}},
  \bibinfo{author}{\bibfnamefont{I.}~\bibnamefont{Iacopini}},
  \bibinfo{author}{\bibfnamefont{V.}~\bibnamefont{Latora}},
  \bibinfo{author}{\bibfnamefont{M.}~\bibnamefont{Lucas}},
  \bibinfo{author}{\bibfnamefont{A.}~\bibnamefont{Patania}},
  \bibinfo{author}{\bibfnamefont{J.-G.} \bibnamefont{Young}}, \bibnamefont{and}
  \bibinfo{author}{\bibfnamefont{G.}~\bibnamefont{Petri}},
  \bibinfo{journal}{Physics Reports} \textbf{\bibinfo{volume}{874}},
  \bibinfo{pages}{1} (\bibinfo{year}{2020}).

\bibitem[{\citenamefont{Bianconi}(2021)}]{bianconi2021higher}
\bibinfo{author}{\bibfnamefont{G.}~\bibnamefont{Bianconi}},
  \emph{\bibinfo{title}{Higher-Order Networks}} (\bibinfo{publisher}{Cambridge
  University Press}, \bibinfo{year}{2021}).

\bibitem[{\citenamefont{Giusti et~al.}(2016)\citenamefont{Giusti, Ghrist, and
  Bassett}}]{giusti2016two}
\bibinfo{author}{\bibfnamefont{C.}~\bibnamefont{Giusti}},
  \bibinfo{author}{\bibfnamefont{R.}~\bibnamefont{Ghrist}}, \bibnamefont{and}
  \bibinfo{author}{\bibfnamefont{D.~S.} \bibnamefont{Bassett}},
  \bibinfo{journal}{Journal of Computational Neuroscience}
  \textbf{\bibinfo{volume}{41}}, \bibinfo{pages}{1} (\bibinfo{year}{2016}).

\bibitem[{\citenamefont{Petri et~al.}(2014)\citenamefont{Petri, Expert,
  Turkheimer, Carhart-Harris, Nutt, Hellyer, and
  Vaccarino}}]{petri2014homological}
\bibinfo{author}{\bibfnamefont{G.}~\bibnamefont{Petri}},
  \bibinfo{author}{\bibfnamefont{P.}~\bibnamefont{Expert}},
  \bibinfo{author}{\bibfnamefont{F.}~\bibnamefont{Turkheimer}},
  \bibinfo{author}{\bibfnamefont{R.}~\bibnamefont{Carhart-Harris}},
  \bibinfo{author}{\bibfnamefont{D.}~\bibnamefont{Nutt}},
  \bibinfo{author}{\bibfnamefont{P.~J.} \bibnamefont{Hellyer}},
  \bibnamefont{and}
  \bibinfo{author}{\bibfnamefont{F.}~\bibnamefont{Vaccarino}},
  \bibinfo{journal}{Journal of The Royal Society Interface}
  \textbf{\bibinfo{volume}{11}}, \bibinfo{pages}{20140873}
  (\bibinfo{year}{2014}).

\bibitem[{\citenamefont{Benson et~al.}(2016)\citenamefont{Benson, Gleich, and
  Leskovec}}]{benson2016higher}
\bibinfo{author}{\bibfnamefont{A.~R.} \bibnamefont{Benson}},
  \bibinfo{author}{\bibfnamefont{D.~F.} \bibnamefont{Gleich}},
  \bibnamefont{and} \bibinfo{author}{\bibfnamefont{J.}~\bibnamefont{Leskovec}},
  \bibinfo{journal}{Science} \textbf{\bibinfo{volume}{353}},
  \bibinfo{pages}{163} (\bibinfo{year}{2016}).

\bibitem[{\citenamefont{Levine et~al.}(2017)\citenamefont{Levine, Bascompte,
  Adler, and Allesina}}]{levine2017beyond}
\bibinfo{author}{\bibfnamefont{J.~M.} \bibnamefont{Levine}},
  \bibinfo{author}{\bibfnamefont{J.}~\bibnamefont{Bascompte}},
  \bibinfo{author}{\bibfnamefont{P.~B.} \bibnamefont{Adler}}, \bibnamefont{and}
  \bibinfo{author}{\bibfnamefont{S.}~\bibnamefont{Allesina}},
  \bibinfo{journal}{Nature} \textbf{\bibinfo{volume}{546}}, \bibinfo{pages}{56}
  (\bibinfo{year}{2017}).

\bibitem[{\citenamefont{Grilli et~al.}(2017)\citenamefont{Grilli, Barab{\'a}s,
  Michalska-Smith, and Allesina}}]{grilli2017higher}
\bibinfo{author}{\bibfnamefont{J.}~\bibnamefont{Grilli}},
  \bibinfo{author}{\bibfnamefont{G.}~\bibnamefont{Barab{\'a}s}},
  \bibinfo{author}{\bibfnamefont{M.~J.} \bibnamefont{Michalska-Smith}},
  \bibnamefont{and} \bibinfo{author}{\bibfnamefont{S.}~\bibnamefont{Allesina}},
  \bibinfo{journal}{Nature} \textbf{\bibinfo{volume}{548}},
  \bibinfo{pages}{210} (\bibinfo{year}{2017}).

\bibitem[{\citenamefont{Lambiotte et~al.}(2019)\citenamefont{Lambiotte,
  Rosvall, and Scholtes}}]{lambiotte2019networks}
\bibinfo{author}{\bibfnamefont{R.}~\bibnamefont{Lambiotte}},
  \bibinfo{author}{\bibfnamefont{M.}~\bibnamefont{Rosvall}}, \bibnamefont{and}
  \bibinfo{author}{\bibfnamefont{I.}~\bibnamefont{Scholtes}},
  \bibinfo{journal}{Nature Physics} \textbf{\bibinfo{volume}{15}},
  \bibinfo{pages}{313} (\bibinfo{year}{2019}).

\bibitem[{\citenamefont{Johnson}(2013)}]{johnson2013hypernetworks}
\bibinfo{author}{\bibfnamefont{J.}~\bibnamefont{Johnson}},
  \emph{\bibinfo{title}{Hypernetworks in the science of complex systems}},
  vol.~\bibinfo{volume}{3} (\bibinfo{publisher}{World Scientific},
  \bibinfo{year}{2013}).

\bibitem[{\citenamefont{Bick et~al.}(2021)\citenamefont{Bick, Gross,
  Harrington, and Schaub}}]{bick2021higher}
\bibinfo{author}{\bibfnamefont{C.}~\bibnamefont{Bick}},
  \bibinfo{author}{\bibfnamefont{E.}~\bibnamefont{Gross}},
  \bibinfo{author}{\bibfnamefont{H.~A.} \bibnamefont{Harrington}},
  \bibnamefont{and} \bibinfo{author}{\bibfnamefont{M.~T.}
  \bibnamefont{Schaub}}, \bibinfo{journal}{arXiv preprint arXiv:2104.11329}
  (\bibinfo{year}{2021}).

\bibitem[{\citenamefont{Battiston et~al.}(2021)\citenamefont{Battiston, Amico,
  Barrat, Bianconi, Ferraz~de Arruda, Franceschiello, Iacopini, K{\'e}fi,
  Latora, Moreno et~al.}}]{battiston2021physics}
\bibinfo{author}{\bibfnamefont{F.}~\bibnamefont{Battiston}},
  \bibinfo{author}{\bibfnamefont{E.}~\bibnamefont{Amico}},
  \bibinfo{author}{\bibfnamefont{A.}~\bibnamefont{Barrat}},
  \bibinfo{author}{\bibfnamefont{G.}~\bibnamefont{Bianconi}},
  \bibinfo{author}{\bibfnamefont{G.}~\bibnamefont{Ferraz~de Arruda}},
  \bibinfo{author}{\bibfnamefont{B.}~\bibnamefont{Franceschiello}},
  \bibinfo{author}{\bibfnamefont{I.}~\bibnamefont{Iacopini}},
  \bibinfo{author}{\bibfnamefont{S.}~\bibnamefont{K{\'e}fi}},
  \bibinfo{author}{\bibfnamefont{V.}~\bibnamefont{Latora}},
  \bibinfo{author}{\bibfnamefont{Y.}~\bibnamefont{Moreno}},
  \bibnamefont{et~al.}, \bibinfo{journal}{Nature Physics}
  \textbf{\bibinfo{volume}{17}}, \bibinfo{pages}{1093} (\bibinfo{year}{2021}).

\bibitem[{\citenamefont{Benson et~al.}(2018)\citenamefont{Benson, Abebe,
  Schaub, Jadbabaie, and Kleinberg}}]{benson2018simplicial}
\bibinfo{author}{\bibfnamefont{A.~R.} \bibnamefont{Benson}},
  \bibinfo{author}{\bibfnamefont{R.}~\bibnamefont{Abebe}},
  \bibinfo{author}{\bibfnamefont{M.~T.} \bibnamefont{Schaub}},
  \bibinfo{author}{\bibfnamefont{A.}~\bibnamefont{Jadbabaie}},
  \bibnamefont{and}
  \bibinfo{author}{\bibfnamefont{J.}~\bibnamefont{Kleinberg}},
  \bibinfo{journal}{Proceedings of the National Academy of Sciences}
  \textbf{\bibinfo{volume}{115}}, \bibinfo{pages}{E11221}
  (\bibinfo{year}{2018}).

\bibitem[{\citenamefont{Liben-Nowell and Kleinberg}(2007)}]{liben2007link}
\bibinfo{author}{\bibfnamefont{D.}~\bibnamefont{Liben-Nowell}}
  \bibnamefont{and}
  \bibinfo{author}{\bibfnamefont{J.}~\bibnamefont{Kleinberg}},
  \bibinfo{journal}{Journal of the American Society for Information Science and
  Technology} \textbf{\bibinfo{volume}{58}}, \bibinfo{pages}{1019}
  (\bibinfo{year}{2007}).

\bibitem[{\citenamefont{Sharma et~al.}(2021)\citenamefont{Sharma, Challa,
  Gupta, and Murty}}]{sharma2021higher}
\bibinfo{author}{\bibfnamefont{G.}~\bibnamefont{Sharma}},
  \bibinfo{author}{\bibfnamefont{A.}~\bibnamefont{Challa}},
  \bibinfo{author}{\bibfnamefont{P.}~\bibnamefont{Gupta}}, \bibnamefont{and}
  \bibinfo{author}{\bibfnamefont{M.~N.} \bibnamefont{Murty}},
  \bibinfo{journal}{arXiv preprint arXiv:2111.00271}  (\bibinfo{year}{2021}).

\bibitem[{\citenamefont{Beentjes and Khamseh}(2020)}]{beentjes2020higher}
\bibinfo{author}{\bibfnamefont{S.~V.} \bibnamefont{Beentjes}} \bibnamefont{and}
  \bibinfo{author}{\bibfnamefont{A.}~\bibnamefont{Khamseh}},
  \bibinfo{journal}{Physical Review E} \textbf{\bibinfo{volume}{102}},
  \bibinfo{pages}{053314} (\bibinfo{year}{2020}).

\bibitem[{\citenamefont{Young et~al.}(2021)\citenamefont{Young, Petri, and
  Peixoto}}]{young2021hypergraph}
\bibinfo{author}{\bibfnamefont{J.-G.} \bibnamefont{Young}},
  \bibinfo{author}{\bibfnamefont{G.}~\bibnamefont{Petri}}, \bibnamefont{and}
  \bibinfo{author}{\bibfnamefont{T.~P.} \bibnamefont{Peixoto}},
  \bibinfo{journal}{Communications Physics} \textbf{\bibinfo{volume}{4}},
  \bibinfo{pages}{1} (\bibinfo{year}{2021}).

\bibitem[{\citenamefont{Musciotto et~al.}(2021)\citenamefont{Musciotto,
  Battiston, and Mantegna}}]{musciotto2021detecting}
\bibinfo{author}{\bibfnamefont{F.}~\bibnamefont{Musciotto}},
  \bibinfo{author}{\bibfnamefont{F.}~\bibnamefont{Battiston}},
  \bibnamefont{and} \bibinfo{author}{\bibfnamefont{R.~N.}
  \bibnamefont{Mantegna}}, \bibinfo{journal}{Communications Physics}
  \textbf{\bibinfo{volume}{4}}, \bibinfo{pages}{1} (\bibinfo{year}{2021}).

\bibitem[{\citenamefont{Sarker et~al.}(2021)\citenamefont{Sarker, Seby, Benson,
  and Jadbabaie}}]{sarker2021higher}
\bibinfo{author}{\bibfnamefont{A.}~\bibnamefont{Sarker}},
  \bibinfo{author}{\bibfnamefont{J.-B.} \bibnamefont{Seby}},
  \bibinfo{author}{\bibfnamefont{A.~R.} \bibnamefont{Benson}},
  \bibnamefont{and}
  \bibinfo{author}{\bibfnamefont{A.}~\bibnamefont{Jadbabaie}},
  \bibinfo{journal}{arXiv preprint arXiv:2108.02091}  (\bibinfo{year}{2021}).

\bibitem[{\citenamefont{Kovalenko
  et~al.}(2021{\natexlab{a}})\citenamefont{Kovalenko, Romance, Aleja,
  Vasilyeva, Criado, Raigorodskii, Flores, Alfaro-Bittner, Perc, and
  Boccaletti}}]{kovalenko2021vector}
\bibinfo{author}{\bibfnamefont{K.}~\bibnamefont{Kovalenko}},
  \bibinfo{author}{\bibfnamefont{M.}~\bibnamefont{Romance}},
  \bibinfo{author}{\bibfnamefont{D.}~\bibnamefont{Aleja}},
  \bibinfo{author}{\bibfnamefont{E.}~\bibnamefont{Vasilyeva}},
  \bibinfo{author}{\bibfnamefont{R.}~\bibnamefont{Criado}},
  \bibinfo{author}{\bibfnamefont{A.~M.} \bibnamefont{Raigorodskii}},
  \bibinfo{author}{\bibfnamefont{J.}~\bibnamefont{Flores}},
  \bibinfo{author}{\bibfnamefont{K.}~\bibnamefont{Alfaro-Bittner}},
  \bibinfo{author}{\bibfnamefont{M.}~\bibnamefont{Perc}}, \bibnamefont{and}
  \bibinfo{author}{\bibfnamefont{S.}~\bibnamefont{Boccaletti}},
  \bibinfo{journal}{arXiv preprint arXiv:2108.13846}
  (\bibinfo{year}{2021}{\natexlab{a}}).

\bibitem[{\citenamefont{Bobrowski and Krioukov}(2021)}]{bobrowski2021random}
\bibinfo{author}{\bibfnamefont{O.}~\bibnamefont{Bobrowski}} \bibnamefont{and}
  \bibinfo{author}{\bibfnamefont{D.}~\bibnamefont{Krioukov}},
  \bibinfo{journal}{arXiv preprint arXiv:2105.12914}  (\bibinfo{year}{2021}).

\bibitem[{\citenamefont{Kovalenko
  et~al.}(2021{\natexlab{b}})\citenamefont{Kovalenko, Sendi{\~n}a-Nadal,
  Khalil, Dainiak, Musatov, Raigorodskii, Alfaro-Bittner, Barzel, and
  Boccaletti}}]{kovalenko2021growing}
\bibinfo{author}{\bibfnamefont{K.}~\bibnamefont{Kovalenko}},
  \bibinfo{author}{\bibfnamefont{I.}~\bibnamefont{Sendi{\~n}a-Nadal}},
  \bibinfo{author}{\bibfnamefont{N.}~\bibnamefont{Khalil}},
  \bibinfo{author}{\bibfnamefont{A.}~\bibnamefont{Dainiak}},
  \bibinfo{author}{\bibfnamefont{D.}~\bibnamefont{Musatov}},
  \bibinfo{author}{\bibfnamefont{A.~M.} \bibnamefont{Raigorodskii}},
  \bibinfo{author}{\bibfnamefont{K.}~\bibnamefont{Alfaro-Bittner}},
  \bibinfo{author}{\bibfnamefont{B.}~\bibnamefont{Barzel}}, \bibnamefont{and}
  \bibinfo{author}{\bibfnamefont{S.}~\bibnamefont{Boccaletti}},
  \bibinfo{journal}{Communications Physics} \textbf{\bibinfo{volume}{4}},
  \bibinfo{pages}{1} (\bibinfo{year}{2021}{\natexlab{b}}).

\bibitem[{\citenamefont{Courtney and Bianconi}(2017)}]{courtney2017weighted}
\bibinfo{author}{\bibfnamefont{O.~T.} \bibnamefont{Courtney}} \bibnamefont{and}
  \bibinfo{author}{\bibfnamefont{G.}~\bibnamefont{Bianconi}},
  \bibinfo{journal}{Physical Review E} \textbf{\bibinfo{volume}{95}},
  \bibinfo{pages}{062301} (\bibinfo{year}{2017}).

\bibitem[{\citenamefont{Courtney and Bianconi}(2018)}]{courtney2018dense}
\bibinfo{author}{\bibfnamefont{O.~T.} \bibnamefont{Courtney}} \bibnamefont{and}
  \bibinfo{author}{\bibfnamefont{G.}~\bibnamefont{Bianconi}},
  \bibinfo{journal}{Physical Review E} \textbf{\bibinfo{volume}{97}},
  \bibinfo{pages}{052303} (\bibinfo{year}{2018}).

\bibitem[{\citenamefont{Bianconi and Rahmede}(2016)}]{bianconi2016network}
\bibinfo{author}{\bibfnamefont{G.}~\bibnamefont{Bianconi}} \bibnamefont{and}
  \bibinfo{author}{\bibfnamefont{C.}~\bibnamefont{Rahmede}},
  \bibinfo{journal}{Physical Review E} \textbf{\bibinfo{volume}{93}},
  \bibinfo{pages}{032315} (\bibinfo{year}{2016}).

\bibitem[{\citenamefont{Mulder and Bianconi}(2018)}]{mulder2018network}
\bibinfo{author}{\bibfnamefont{D.}~\bibnamefont{Mulder}} \bibnamefont{and}
  \bibinfo{author}{\bibfnamefont{G.}~\bibnamefont{Bianconi}},
  \bibinfo{journal}{Journal of Statistical Physics}
  \textbf{\bibinfo{volume}{173}}, \bibinfo{pages}{783} (\bibinfo{year}{2018}).

\bibitem[{\citenamefont{Petri and Barrat}(2018)}]{petri2018simplicial}
\bibinfo{author}{\bibfnamefont{G.}~\bibnamefont{Petri}} \bibnamefont{and}
  \bibinfo{author}{\bibfnamefont{A.}~\bibnamefont{Barrat}},
  \bibinfo{journal}{Physical Review Letters} \textbf{\bibinfo{volume}{121}},
  \bibinfo{pages}{228301} (\bibinfo{year}{2018}).

\bibitem[{\citenamefont{Young et~al.}(2017)\citenamefont{Young, Petri,
  Vaccarino, and Patania}}]{young2017construction}
\bibinfo{author}{\bibfnamefont{J.-G.} \bibnamefont{Young}},
  \bibinfo{author}{\bibfnamefont{G.}~\bibnamefont{Petri}},
  \bibinfo{author}{\bibfnamefont{F.}~\bibnamefont{Vaccarino}},
  \bibnamefont{and} \bibinfo{author}{\bibfnamefont{A.}~\bibnamefont{Patania}},
  \bibinfo{journal}{Physical Review E} \textbf{\bibinfo{volume}{96}},
  \bibinfo{pages}{032312} (\bibinfo{year}{2017}).

\bibitem[{\citenamefont{Courtney and Bianconi}(2016)}]{courtney2016generalized}
\bibinfo{author}{\bibfnamefont{O.~T.} \bibnamefont{Courtney}} \bibnamefont{and}
  \bibinfo{author}{\bibfnamefont{G.}~\bibnamefont{Bianconi}},
  \bibinfo{journal}{Physical Review E} \textbf{\bibinfo{volume}{93}},
  \bibinfo{pages}{062311} (\bibinfo{year}{2016}).

\bibitem[{\citenamefont{Yen}(2021)}]{yen2021construction}
\bibinfo{author}{\bibfnamefont{T.-C.} \bibnamefont{Yen}},
  \bibinfo{journal}{Physical Review E} \textbf{\bibinfo{volume}{104}},
  \bibinfo{pages}{L042303} (\bibinfo{year}{2021}).

\bibitem[{\citenamefont{Chodrow}(2020)}]{chodrow2020configuration}
\bibinfo{author}{\bibfnamefont{P.~S.} \bibnamefont{Chodrow}},
  \bibinfo{journal}{Journal of Complex Networks} \textbf{\bibinfo{volume}{8}},
  \bibinfo{pages}{cnaa018} (\bibinfo{year}{2020}).

\bibitem[{\citenamefont{Chodrow and Mellor}(2020)}]{chodrow2020annotated}
\bibinfo{author}{\bibfnamefont{P.}~\bibnamefont{Chodrow}} \bibnamefont{and}
  \bibinfo{author}{\bibfnamefont{A.}~\bibnamefont{Mellor}},
  \bibinfo{journal}{Applied Network Science} \textbf{\bibinfo{volume}{5}},
  \bibinfo{pages}{1} (\bibinfo{year}{2020}).

\bibitem[{\citenamefont{Zhang et~al.}(2021{\natexlab{a}})\citenamefont{Zhang,
  Garas, and Scholtes}}]{zhang2021higher}
\bibinfo{author}{\bibfnamefont{Y.}~\bibnamefont{Zhang}},
  \bibinfo{author}{\bibfnamefont{A.}~\bibnamefont{Garas}}, \bibnamefont{and}
  \bibinfo{author}{\bibfnamefont{I.}~\bibnamefont{Scholtes}},
  \bibinfo{journal}{Journal of Physics: Complexity}
  \textbf{\bibinfo{volume}{2}}, \bibinfo{pages}{015007}
  (\bibinfo{year}{2021}{\natexlab{a}}).

\bibitem[{\citenamefont{Vasilyeva et~al.}(2021)\citenamefont{Vasilyeva, Kozlov,
  Alfaro-Bittner, Musatov, Raigorodskii, Perc, and
  Boccaletti}}]{vasilyeva2021multilayer}
\bibinfo{author}{\bibfnamefont{E.}~\bibnamefont{Vasilyeva}},
  \bibinfo{author}{\bibfnamefont{A.}~\bibnamefont{Kozlov}},
  \bibinfo{author}{\bibfnamefont{K.}~\bibnamefont{Alfaro-Bittner}},
  \bibinfo{author}{\bibfnamefont{D.}~\bibnamefont{Musatov}},
  \bibinfo{author}{\bibfnamefont{A.}~\bibnamefont{Raigorodskii}},
  \bibinfo{author}{\bibfnamefont{M.}~\bibnamefont{Perc}}, \bibnamefont{and}
  \bibinfo{author}{\bibfnamefont{S.}~\bibnamefont{Boccaletti}},
  \bibinfo{journal}{Scientific Reports} \textbf{\bibinfo{volume}{11}},
  \bibinfo{pages}{1} (\bibinfo{year}{2021}).

\bibitem[{\citenamefont{Cencetti et~al.}(2021)\citenamefont{Cencetti,
  Battiston, Lepri, and Karsai}}]{cencetti2021temporal}
\bibinfo{author}{\bibfnamefont{G.}~\bibnamefont{Cencetti}},
  \bibinfo{author}{\bibfnamefont{F.}~\bibnamefont{Battiston}},
  \bibinfo{author}{\bibfnamefont{B.}~\bibnamefont{Lepri}}, \bibnamefont{and}
  \bibinfo{author}{\bibfnamefont{M.}~\bibnamefont{Karsai}},
  \bibinfo{journal}{Scientific Reports} \textbf{\bibinfo{volume}{11}},
  \bibinfo{pages}{1} (\bibinfo{year}{2021}).

\bibitem[{\citenamefont{Yin et~al.}(2018)\citenamefont{Yin, Benson, and
  Leskovec}}]{yin2018higher}
\bibinfo{author}{\bibfnamefont{H.}~\bibnamefont{Yin}},
  \bibinfo{author}{\bibfnamefont{A.~R.} \bibnamefont{Benson}},
  \bibnamefont{and} \bibinfo{author}{\bibfnamefont{J.}~\bibnamefont{Leskovec}},
  \bibinfo{journal}{Physical Review E} \textbf{\bibinfo{volume}{97}},
  \bibinfo{pages}{052306} (\bibinfo{year}{2018}).

\bibitem[{\citenamefont{Amburg et~al.}(2020)\citenamefont{Amburg, Veldt, and
  Benson}}]{amburg2020clustering}
\bibinfo{author}{\bibfnamefont{I.}~\bibnamefont{Amburg}},
  \bibinfo{author}{\bibfnamefont{N.}~\bibnamefont{Veldt}}, \bibnamefont{and}
  \bibinfo{author}{\bibfnamefont{A.}~\bibnamefont{Benson}}, in
  \emph{\bibinfo{booktitle}{Proceedings of The Web Conference 2020}}
  (\bibinfo{year}{2020}), pp. \bibinfo{pages}{706--717}.

\bibitem[{\citenamefont{Krishnagopal and
  Bianconi}(2021)}]{krishnagopal2021spectral}
\bibinfo{author}{\bibfnamefont{S.}~\bibnamefont{Krishnagopal}}
  \bibnamefont{and} \bibinfo{author}{\bibfnamefont{G.}~\bibnamefont{Bianconi}},
  \bibinfo{journal}{Physical Review E} \textbf{\bibinfo{volume}{104}},
  \bibinfo{pages}{064303} (\bibinfo{year}{2021}).

\bibitem[{\citenamefont{Chodrow et~al.}(2021)\citenamefont{Chodrow, Veldt, and
  Benson}}]{chodrow2021generative}
\bibinfo{author}{\bibfnamefont{P.~S.} \bibnamefont{Chodrow}},
  \bibinfo{author}{\bibfnamefont{N.}~\bibnamefont{Veldt}}, \bibnamefont{and}
  \bibinfo{author}{\bibfnamefont{A.~R.} \bibnamefont{Benson}},
  \bibinfo{journal}{arXiv preprint arXiv:2101.09611}  (\bibinfo{year}{2021}).

\bibitem[{\citenamefont{Tudisco and Higham}(2021)}]{tudisco2021node}
\bibinfo{author}{\bibfnamefont{F.}~\bibnamefont{Tudisco}} \bibnamefont{and}
  \bibinfo{author}{\bibfnamefont{D.~J.} \bibnamefont{Higham}},
  \bibinfo{journal}{Communications Physics} \textbf{\bibinfo{volume}{4}},
  \bibinfo{pages}{1} (\bibinfo{year}{2021}).

\bibitem[{\citenamefont{Veldt et~al.}(2021)\citenamefont{Veldt, Benson, and
  Kleinberg}}]{veldt2021higher}
\bibinfo{author}{\bibfnamefont{N.}~\bibnamefont{Veldt}},
  \bibinfo{author}{\bibfnamefont{A.~R.} \bibnamefont{Benson}},
  \bibnamefont{and}
  \bibinfo{author}{\bibfnamefont{J.}~\bibnamefont{Kleinberg}},
  \bibinfo{journal}{arXiv preprint arXiv:2103.11818}  (\bibinfo{year}{2021}).

\bibitem[{\citenamefont{Sun and Bianconi}(2021)}]{sun2021higher}
\bibinfo{author}{\bibfnamefont{H.}~\bibnamefont{Sun}} \bibnamefont{and}
  \bibinfo{author}{\bibfnamefont{G.}~\bibnamefont{Bianconi}},
  \bibinfo{journal}{arXiv preprint arXiv:2104.05457}  (\bibinfo{year}{2021}).

\bibitem[{\citenamefont{Bianconi et~al.}(2019)\citenamefont{Bianconi, Kryven,
  and Ziff}}]{bianconi2019percolation}
\bibinfo{author}{\bibfnamefont{G.}~\bibnamefont{Bianconi}},
  \bibinfo{author}{\bibfnamefont{I.}~\bibnamefont{Kryven}}, \bibnamefont{and}
  \bibinfo{author}{\bibfnamefont{R.~M.} \bibnamefont{Ziff}},
  \bibinfo{journal}{Physical Review E} \textbf{\bibinfo{volume}{100}},
  \bibinfo{pages}{062311} (\bibinfo{year}{2019}).

\bibitem[{\citenamefont{Lee et~al.}(2021)\citenamefont{Lee, Lee, Oh, Lee, and
  Kahng}}]{lee2021homological}
\bibinfo{author}{\bibfnamefont{Y.}~\bibnamefont{Lee}},
  \bibinfo{author}{\bibfnamefont{J.}~\bibnamefont{Lee}},
  \bibinfo{author}{\bibfnamefont{S.~M.} \bibnamefont{Oh}},
  \bibinfo{author}{\bibfnamefont{D.}~\bibnamefont{Lee}}, \bibnamefont{and}
  \bibinfo{author}{\bibfnamefont{B.}~\bibnamefont{Kahng}},
  \bibinfo{journal}{Chaos: An Interdisciplinary Journal of Nonlinear Science}
  \textbf{\bibinfo{volume}{31}}, \bibinfo{pages}{041102}
  (\bibinfo{year}{2021}).

\bibitem[{\citenamefont{Landry and Restrepo}(2021)}]{landry2021hypergraph}
\bibinfo{author}{\bibfnamefont{N.~W.} \bibnamefont{Landry}} \bibnamefont{and}
  \bibinfo{author}{\bibfnamefont{J.~G.} \bibnamefont{Restrepo}},
  \bibinfo{journal}{arXiv preprint arXiv:2109.01099}  (\bibinfo{year}{2021}).

\bibitem[{\citenamefont{Bairey et~al.}(2016)\citenamefont{Bairey, Kelsic, and
  Kishony}}]{bairey2016high}
\bibinfo{author}{\bibfnamefont{E.}~\bibnamefont{Bairey}},
  \bibinfo{author}{\bibfnamefont{E.~D.} \bibnamefont{Kelsic}},
  \bibnamefont{and} \bibinfo{author}{\bibfnamefont{R.}~\bibnamefont{Kishony}},
  \bibinfo{journal}{Nature Communications} \textbf{\bibinfo{volume}{7}},
  \bibinfo{pages}{1} (\bibinfo{year}{2016}).

\bibitem[{\citenamefont{Pikovsky et~al.}(2003)\citenamefont{Pikovsky,
  Rosenblum, and Kurths}}]{pikovsky2003synchronization}
\bibinfo{author}{\bibfnamefont{A.}~\bibnamefont{Pikovsky}},
  \bibinfo{author}{\bibfnamefont{M.}~\bibnamefont{Rosenblum}},
  \bibnamefont{and} \bibinfo{author}{\bibfnamefont{J.}~\bibnamefont{Kurths}},
  \emph{\bibinfo{title}{Synchronization: a universal concept in nonlinear
  sciences}}, vol.~\bibinfo{volume}{12} (\bibinfo{publisher}{Cambridge
  University Press}, \bibinfo{year}{2003}).

\bibitem[{\citenamefont{Boccaletti et~al.}(2002)\citenamefont{Boccaletti,
  Kurths, Osipov, Valladares, and Zhou}}]{boccaletti2002synchronization}
\bibinfo{author}{\bibfnamefont{S.}~\bibnamefont{Boccaletti}},
  \bibinfo{author}{\bibfnamefont{J.}~\bibnamefont{Kurths}},
  \bibinfo{author}{\bibfnamefont{G.}~\bibnamefont{Osipov}},
  \bibinfo{author}{\bibfnamefont{D.}~\bibnamefont{Valladares}},
  \bibnamefont{and} \bibinfo{author}{\bibfnamefont{C.}~\bibnamefont{Zhou}},
  \bibinfo{journal}{Physics Reports} \textbf{\bibinfo{volume}{366}},
  \bibinfo{pages}{1} (\bibinfo{year}{2002}).

\bibitem[{\citenamefont{Wang et~al.}(2009)\citenamefont{Wang, Perc, Duan, and
  Chen}}]{wang2009synchronization}
\bibinfo{author}{\bibfnamefont{Q.}~\bibnamefont{Wang}},
  \bibinfo{author}{\bibfnamefont{M.}~\bibnamefont{Perc}},
  \bibinfo{author}{\bibfnamefont{Z.}~\bibnamefont{Duan}}, \bibnamefont{and}
  \bibinfo{author}{\bibfnamefont{G.}~\bibnamefont{Chen}},
  \bibinfo{journal}{Physical Review E} \textbf{\bibinfo{volume}{80}},
  \bibinfo{pages}{026206} (\bibinfo{year}{2009}).

\bibitem[{\citenamefont{Majhi et~al.}(2019{\natexlab{a}})\citenamefont{Majhi,
  Ghosh, and Kurths}}]{majhi2019emergence}
\bibinfo{author}{\bibfnamefont{S.}~\bibnamefont{Majhi}},
  \bibinfo{author}{\bibfnamefont{D.}~\bibnamefont{Ghosh}}, \bibnamefont{and}
  \bibinfo{author}{\bibfnamefont{J.}~\bibnamefont{Kurths}},
  \bibinfo{journal}{Physical Review E} \textbf{\bibinfo{volume}{99}},
  \bibinfo{pages}{012308} (\bibinfo{year}{2019}{\natexlab{a}}).

\bibitem[{\citenamefont{Arenas et~al.}(2008)\citenamefont{Arenas,
  D{\'\i}az-Guilera, Kurths, Moreno, and Zhou}}]{arenas2008synchronization}
\bibinfo{author}{\bibfnamefont{A.}~\bibnamefont{Arenas}},
  \bibinfo{author}{\bibfnamefont{A.}~\bibnamefont{D{\'\i}az-Guilera}},
  \bibinfo{author}{\bibfnamefont{J.}~\bibnamefont{Kurths}},
  \bibinfo{author}{\bibfnamefont{Y.}~\bibnamefont{Moreno}}, \bibnamefont{and}
  \bibinfo{author}{\bibfnamefont{C.}~\bibnamefont{Zhou}},
  \bibinfo{journal}{Physics Reports} \textbf{\bibinfo{volume}{469}},
  \bibinfo{pages}{93} (\bibinfo{year}{2008}).

\bibitem[{\citenamefont{Majhi and Ghosh}(2017)}]{majhi2017synchronization}
\bibinfo{author}{\bibfnamefont{S.}~\bibnamefont{Majhi}} \bibnamefont{and}
  \bibinfo{author}{\bibfnamefont{D.}~\bibnamefont{Ghosh}},
  \bibinfo{journal}{Chaos: An Interdisciplinary Journal of Nonlinear Science}
  \textbf{\bibinfo{volume}{27}}, \bibinfo{pages}{053115}
  (\bibinfo{year}{2017}).

\bibitem[{\citenamefont{Tanaka and Aoyagi}(2011)}]{tanaka2011multistable}
\bibinfo{author}{\bibfnamefont{T.}~\bibnamefont{Tanaka}} \bibnamefont{and}
  \bibinfo{author}{\bibfnamefont{T.}~\bibnamefont{Aoyagi}},
  \bibinfo{journal}{Physical Review Letters} \textbf{\bibinfo{volume}{106}},
  \bibinfo{pages}{224101} (\bibinfo{year}{2011}).

\bibitem[{\citenamefont{Skardal and Arenas}(2019)}]{skardal2019abrupt}
\bibinfo{author}{\bibfnamefont{P.~S.} \bibnamefont{Skardal}} \bibnamefont{and}
  \bibinfo{author}{\bibfnamefont{A.}~\bibnamefont{Arenas}},
  \bibinfo{journal}{Physical Review Letters} \textbf{\bibinfo{volume}{122}},
  \bibinfo{pages}{248301} (\bibinfo{year}{2019}).

\bibitem[{\citenamefont{Skardal and Arenas}(2020)}]{skardal2020higher}
\bibinfo{author}{\bibfnamefont{P.~S.} \bibnamefont{Skardal}} \bibnamefont{and}
  \bibinfo{author}{\bibfnamefont{A.}~\bibnamefont{Arenas}},
  \bibinfo{journal}{Communications Physics} \textbf{\bibinfo{volume}{3}},
  \bibinfo{pages}{1} (\bibinfo{year}{2020}).

\bibitem[{\citenamefont{Gambuzza et~al.}(2021)\citenamefont{Gambuzza, Di~Patti,
  Gallo, Lepri, Romance, Criado, Frasca, Latora, and
  Boccaletti}}]{gambuzza2021stability}
\bibinfo{author}{\bibfnamefont{L.~V.} \bibnamefont{Gambuzza}},
  \bibinfo{author}{\bibfnamefont{F.}~\bibnamefont{Di~Patti}},
  \bibinfo{author}{\bibfnamefont{L.}~\bibnamefont{Gallo}},
  \bibinfo{author}{\bibfnamefont{S.}~\bibnamefont{Lepri}},
  \bibinfo{author}{\bibfnamefont{M.}~\bibnamefont{Romance}},
  \bibinfo{author}{\bibfnamefont{R.}~\bibnamefont{Criado}},
  \bibinfo{author}{\bibfnamefont{M.}~\bibnamefont{Frasca}},
  \bibinfo{author}{\bibfnamefont{V.}~\bibnamefont{Latora}}, \bibnamefont{and}
  \bibinfo{author}{\bibfnamefont{S.}~\bibnamefont{Boccaletti}},
  \bibinfo{journal}{Nature Communications} \textbf{\bibinfo{volume}{12}},
  \bibinfo{pages}{1} (\bibinfo{year}{2021}).

\bibitem[{\citenamefont{Dai et~al.}(2021)\citenamefont{Dai, Kovalenko, Molodyk,
  Wang, Li, Musatov, Raigorodskii, Alfaro-Bittner, Cooper, Bianconi
  et~al.}}]{dai2021d}
\bibinfo{author}{\bibfnamefont{X.}~\bibnamefont{Dai}},
  \bibinfo{author}{\bibfnamefont{K.}~\bibnamefont{Kovalenko}},
  \bibinfo{author}{\bibfnamefont{M.}~\bibnamefont{Molodyk}},
  \bibinfo{author}{\bibfnamefont{Z.}~\bibnamefont{Wang}},
  \bibinfo{author}{\bibfnamefont{X.}~\bibnamefont{Li}},
  \bibinfo{author}{\bibfnamefont{D.}~\bibnamefont{Musatov}},
  \bibinfo{author}{\bibfnamefont{A.}~\bibnamefont{Raigorodskii}},
  \bibinfo{author}{\bibfnamefont{K.}~\bibnamefont{Alfaro-Bittner}},
  \bibinfo{author}{\bibfnamefont{G.}~\bibnamefont{Cooper}},
  \bibinfo{author}{\bibfnamefont{G.}~\bibnamefont{Bianconi}},
  \bibnamefont{et~al.}, \bibinfo{journal}{Chaos, Solitons \& Fractals}
  \textbf{\bibinfo{volume}{146}}, \bibinfo{pages}{110888}
  (\bibinfo{year}{2021}).

\bibitem[{\citenamefont{Chandra et~al.}(2019)\citenamefont{Chandra, Girvan, and
  Ott}}]{chandra2019continuous}
\bibinfo{author}{\bibfnamefont{S.}~\bibnamefont{Chandra}},
  \bibinfo{author}{\bibfnamefont{M.}~\bibnamefont{Girvan}}, \bibnamefont{and}
  \bibinfo{author}{\bibfnamefont{E.}~\bibnamefont{Ott}},
  \bibinfo{journal}{Physical Review X} \textbf{\bibinfo{volume}{9}},
  \bibinfo{pages}{011002} (\bibinfo{year}{2019}).

\bibitem[{\citenamefont{Kovalenko
  et~al.}(2021{\natexlab{c}})\citenamefont{Kovalenko, Dai, Alfaro-Bittner,
  Raigorodskii, Perc, and Boccaletti}}]{kovalenko2021contrarians}
\bibinfo{author}{\bibfnamefont{K.}~\bibnamefont{Kovalenko}},
  \bibinfo{author}{\bibfnamefont{X.}~\bibnamefont{Dai}},
  \bibinfo{author}{\bibfnamefont{K.}~\bibnamefont{Alfaro-Bittner}},
  \bibinfo{author}{\bibfnamefont{A.}~\bibnamefont{Raigorodskii}},
  \bibinfo{author}{\bibfnamefont{M.}~\bibnamefont{Perc}}, \bibnamefont{and}
  \bibinfo{author}{\bibfnamefont{S.}~\bibnamefont{Boccaletti}},
  \bibinfo{journal}{Physical Review Letters} \textbf{\bibinfo{volume}{127}},
  \bibinfo{pages}{258301} (\bibinfo{year}{2021}{\natexlab{c}}).

\bibitem[{\citenamefont{Mill{\'a}n et~al.}(2020)\citenamefont{Mill{\'a}n,
  Torres, and Bianconi}}]{millan2020explosive}
\bibinfo{author}{\bibfnamefont{A.~P.} \bibnamefont{Mill{\'a}n}},
  \bibinfo{author}{\bibfnamefont{J.~J.} \bibnamefont{Torres}},
  \bibnamefont{and} \bibinfo{author}{\bibfnamefont{G.}~\bibnamefont{Bianconi}},
  \bibinfo{journal}{Physical Review Letters} \textbf{\bibinfo{volume}{124}},
  \bibinfo{pages}{218301} (\bibinfo{year}{2020}).

\bibitem[{\citenamefont{Ghorbanchian et~al.}(2021)\citenamefont{Ghorbanchian,
  Restrepo, Torres, and Bianconi}}]{ghorbanchian2021higher}
\bibinfo{author}{\bibfnamefont{R.}~\bibnamefont{Ghorbanchian}},
  \bibinfo{author}{\bibfnamefont{J.~G.} \bibnamefont{Restrepo}},
  \bibinfo{author}{\bibfnamefont{J.~J.} \bibnamefont{Torres}},
  \bibnamefont{and} \bibinfo{author}{\bibfnamefont{G.}~\bibnamefont{Bianconi}},
  \bibinfo{journal}{Communications Physics} \textbf{\bibinfo{volume}{4}},
  \bibinfo{pages}{1} (\bibinfo{year}{2021}).

\bibitem[{\citenamefont{Chutani et~al.}(2021)\citenamefont{Chutani, Tadic, and
  Gupte}}]{chutani2021hysteresis}
\bibinfo{author}{\bibfnamefont{M.}~\bibnamefont{Chutani}},
  \bibinfo{author}{\bibfnamefont{B.}~\bibnamefont{Tadic}}, \bibnamefont{and}
  \bibinfo{author}{\bibfnamefont{N.}~\bibnamefont{Gupte}},
  \bibinfo{journal}{arXiv preprint arXiv:2106.04355}  (\bibinfo{year}{2021}).

\bibitem[{\citenamefont{Kachhvah and Jalan}(2021)}]{kachhvah2021hebbian}
\bibinfo{author}{\bibfnamefont{A.~D.} \bibnamefont{Kachhvah}} \bibnamefont{and}
  \bibinfo{author}{\bibfnamefont{S.}~\bibnamefont{Jalan}},
  \bibinfo{journal}{arXiv preprint arXiv:2110.13553}  (\bibinfo{year}{2021}).

\bibitem[{\citenamefont{Arnaudon et~al.}(2021)\citenamefont{Arnaudon, Peach,
  Petri, and Expert}}]{arnaudon2021connecting}
\bibinfo{author}{\bibfnamefont{A.}~\bibnamefont{Arnaudon}},
  \bibinfo{author}{\bibfnamefont{R.~L.} \bibnamefont{Peach}},
  \bibinfo{author}{\bibfnamefont{G.}~\bibnamefont{Petri}}, \bibnamefont{and}
  \bibinfo{author}{\bibfnamefont{P.}~\bibnamefont{Expert}},
  \bibinfo{journal}{arXiv preprint arXiv:2111.11073}  (\bibinfo{year}{2021}).

\bibitem[{\citenamefont{Lucas et~al.}(2020)\citenamefont{Lucas, Cencetti, and
  Battiston}}]{lucas2020multiorder}
\bibinfo{author}{\bibfnamefont{M.}~\bibnamefont{Lucas}},
  \bibinfo{author}{\bibfnamefont{G.}~\bibnamefont{Cencetti}}, \bibnamefont{and}
  \bibinfo{author}{\bibfnamefont{F.}~\bibnamefont{Battiston}},
  \bibinfo{journal}{Physical Review Research} \textbf{\bibinfo{volume}{2}},
  \bibinfo{pages}{033410} (\bibinfo{year}{2020}).

\bibitem[{\citenamefont{Majhi et~al.}(2020)\citenamefont{Majhi, Chowdhury, and
  Ghosh}}]{majhi2020perspective}
\bibinfo{author}{\bibfnamefont{S.}~\bibnamefont{Majhi}},
  \bibinfo{author}{\bibfnamefont{S.~N.} \bibnamefont{Chowdhury}},
  \bibnamefont{and} \bibinfo{author}{\bibfnamefont{D.}~\bibnamefont{Ghosh}},
  \bibinfo{journal}{EPL (Europhysics Letters)} \textbf{\bibinfo{volume}{132}},
  \bibinfo{pages}{20001} (\bibinfo{year}{2020}).

\bibitem[{\citenamefont{Skardal et~al.}(2021)\citenamefont{Skardal,
  Arola-Fern{\'a}ndez, Taylor, and Arenas}}]{skardal2021higher}
\bibinfo{author}{\bibfnamefont{P.~S.} \bibnamefont{Skardal}},
  \bibinfo{author}{\bibfnamefont{L.}~\bibnamefont{Arola-Fern{\'a}ndez}},
  \bibinfo{author}{\bibfnamefont{D.}~\bibnamefont{Taylor}}, \bibnamefont{and}
  \bibinfo{author}{\bibfnamefont{A.}~\bibnamefont{Arenas}},
  \bibinfo{journal}{Physical Review Research} \textbf{\bibinfo{volume}{3}},
  \bibinfo{pages}{043193} (\bibinfo{year}{2021}).

\bibitem[{\citenamefont{Carletti
  et~al.}(2020{\natexlab{a}})\citenamefont{Carletti, Fanelli, and
  Nicoletti}}]{carletti2020dynamical}
\bibinfo{author}{\bibfnamefont{T.}~\bibnamefont{Carletti}},
  \bibinfo{author}{\bibfnamefont{D.}~\bibnamefont{Fanelli}}, \bibnamefont{and}
  \bibinfo{author}{\bibfnamefont{S.}~\bibnamefont{Nicoletti}},
  \bibinfo{journal}{Journal of Physics: Complexity}
  \textbf{\bibinfo{volume}{1}}, \bibinfo{pages}{035006}
  (\bibinfo{year}{2020}{\natexlab{a}}).

\bibitem[{\citenamefont{Zhang et~al.}(2021{\natexlab{b}})\citenamefont{Zhang,
  Latora, and Motter}}]{zhang2021unified}
\bibinfo{author}{\bibfnamefont{Y.}~\bibnamefont{Zhang}},
  \bibinfo{author}{\bibfnamefont{V.}~\bibnamefont{Latora}}, \bibnamefont{and}
  \bibinfo{author}{\bibfnamefont{A.~E.} \bibnamefont{Motter}},
  \bibinfo{journal}{Communications Physics} \textbf{\bibinfo{volume}{4}},
  \bibinfo{pages}{1} (\bibinfo{year}{2021}{\natexlab{b}}).

\bibitem[{\citenamefont{Mulas et~al.}(2020)\citenamefont{Mulas, Kuehn, and
  Jost}}]{mulas2020coupled}
\bibinfo{author}{\bibfnamefont{R.}~\bibnamefont{Mulas}},
  \bibinfo{author}{\bibfnamefont{C.}~\bibnamefont{Kuehn}}, \bibnamefont{and}
  \bibinfo{author}{\bibfnamefont{J.}~\bibnamefont{Jost}},
  \bibinfo{journal}{Physical Review E} \textbf{\bibinfo{volume}{101}},
  \bibinfo{pages}{062313} (\bibinfo{year}{2020}).

\bibitem[{\citenamefont{B{\"o}hle et~al.}(2021)\citenamefont{B{\"o}hle, Kuehn,
  Mulas, and Jost}}]{bohle2021coupled}
\bibinfo{author}{\bibfnamefont{T.}~\bibnamefont{B{\"o}hle}},
  \bibinfo{author}{\bibfnamefont{C.}~\bibnamefont{Kuehn}},
  \bibinfo{author}{\bibfnamefont{R.}~\bibnamefont{Mulas}}, \bibnamefont{and}
  \bibinfo{author}{\bibfnamefont{J.}~\bibnamefont{Jost}},
  \bibinfo{journal}{arXiv preprint arXiv:2102.02272}  (\bibinfo{year}{2021}).

\bibitem[{\citenamefont{Parastesh et~al.}(2022)\citenamefont{Parastesh,
  Mehrabbeik, Rajagopal, Jafari, and Perc}}]{parastesh2022synchronization}
\bibinfo{author}{\bibfnamefont{F.}~\bibnamefont{Parastesh}},
  \bibinfo{author}{\bibfnamefont{M.}~\bibnamefont{Mehrabbeik}},
  \bibinfo{author}{\bibfnamefont{K.}~\bibnamefont{Rajagopal}},
  \bibinfo{author}{\bibfnamefont{S.}~\bibnamefont{Jafari}}, \bibnamefont{and}
  \bibinfo{author}{\bibfnamefont{M.}~\bibnamefont{Perc}},
  \bibinfo{journal}{Chaos: An Interdisciplinary Journal of Nonlinear Science}
  \textbf{\bibinfo{volume}{32}}, \bibinfo{pages}{013125}
  (\bibinfo{year}{2022}).

\bibitem[{\citenamefont{Castellano et~al.}(2009)\citenamefont{Castellano,
  Fortunato, and Loreto}}]{castellano2009statistical}
\bibinfo{author}{\bibfnamefont{C.}~\bibnamefont{Castellano}},
  \bibinfo{author}{\bibfnamefont{S.}~\bibnamefont{Fortunato}},
  \bibnamefont{and} \bibinfo{author}{\bibfnamefont{V.}~\bibnamefont{Loreto}},
  \bibinfo{journal}{Reviews of Modern Physics} \textbf{\bibinfo{volume}{81}},
  \bibinfo{pages}{591} (\bibinfo{year}{2009}).

\bibitem[{\citenamefont{Malmgren et~al.}(2009)\citenamefont{Malmgren, Stouffer,
  Campanharo, and Amaral}}]{malmgren2009universality}
\bibinfo{author}{\bibfnamefont{R.~D.} \bibnamefont{Malmgren}},
  \bibinfo{author}{\bibfnamefont{D.~B.} \bibnamefont{Stouffer}},
  \bibinfo{author}{\bibfnamefont{A.~S.} \bibnamefont{Campanharo}},
  \bibnamefont{and} \bibinfo{author}{\bibfnamefont{L.~A.~N.}
  \bibnamefont{Amaral}}, \bibinfo{journal}{Science}
  \textbf{\bibinfo{volume}{325}}, \bibinfo{pages}{1696} (\bibinfo{year}{2009}).

\bibitem[{\citenamefont{Iacopini et~al.}(2019)\citenamefont{Iacopini, Petri,
  Barrat, and Latora}}]{iacopini2019simplicial}
\bibinfo{author}{\bibfnamefont{I.}~\bibnamefont{Iacopini}},
  \bibinfo{author}{\bibfnamefont{G.}~\bibnamefont{Petri}},
  \bibinfo{author}{\bibfnamefont{A.}~\bibnamefont{Barrat}}, \bibnamefont{and}
  \bibinfo{author}{\bibfnamefont{V.}~\bibnamefont{Latora}},
  \bibinfo{journal}{Nature communications} \textbf{\bibinfo{volume}{10}},
  \bibinfo{pages}{1} (\bibinfo{year}{2019}).

\bibitem[{\citenamefont{Chowdhary et~al.}(2021)\citenamefont{Chowdhary, Kumar,
  Cencetti, Iacopini, and Battiston}}]{chowdhary2021simplicial}
\bibinfo{author}{\bibfnamefont{S.}~\bibnamefont{Chowdhary}},
  \bibinfo{author}{\bibfnamefont{A.}~\bibnamefont{Kumar}},
  \bibinfo{author}{\bibfnamefont{G.}~\bibnamefont{Cencetti}},
  \bibinfo{author}{\bibfnamefont{I.}~\bibnamefont{Iacopini}}, \bibnamefont{and}
  \bibinfo{author}{\bibfnamefont{F.}~\bibnamefont{Battiston}},
  \bibinfo{journal}{arXiv preprint arXiv:2105.04455}  (\bibinfo{year}{2021}).

\bibitem[{\citenamefont{de~Arruda et~al.}(2020)\citenamefont{de~Arruda, Petri,
  and Moreno}}]{de2020social}
\bibinfo{author}{\bibfnamefont{G.~F.} \bibnamefont{de~Arruda}},
  \bibinfo{author}{\bibfnamefont{G.}~\bibnamefont{Petri}}, \bibnamefont{and}
  \bibinfo{author}{\bibfnamefont{Y.}~\bibnamefont{Moreno}},
  \bibinfo{journal}{Physical Review Research} \textbf{\bibinfo{volume}{2}},
  \bibinfo{pages}{023032} (\bibinfo{year}{2020}).

\bibitem[{\citenamefont{Ritchie et~al.}(2014)\citenamefont{Ritchie, Berthouze,
  House, and Kiss}}]{ritchie2014higher}
\bibinfo{author}{\bibfnamefont{M.}~\bibnamefont{Ritchie}},
  \bibinfo{author}{\bibfnamefont{L.}~\bibnamefont{Berthouze}},
  \bibinfo{author}{\bibfnamefont{T.}~\bibnamefont{House}}, \bibnamefont{and}
  \bibinfo{author}{\bibfnamefont{I.~Z.} \bibnamefont{Kiss}},
  \bibinfo{journal}{Journal of Theoretical Biology}
  \textbf{\bibinfo{volume}{348}}, \bibinfo{pages}{21} (\bibinfo{year}{2014}).

\bibitem[{\citenamefont{Landry and Restrepo}(2020)}]{landry2020effect}
\bibinfo{author}{\bibfnamefont{N.~W.} \bibnamefont{Landry}} \bibnamefont{and}
  \bibinfo{author}{\bibfnamefont{J.~G.} \bibnamefont{Restrepo}},
  \bibinfo{journal}{Chaos: An Interdisciplinary Journal of Nonlinear Science}
  \textbf{\bibinfo{volume}{30}}, \bibinfo{pages}{103117}
  (\bibinfo{year}{2020}).

\bibitem[{\citenamefont{St-Onge
  et~al.}(2021{\natexlab{a}})\citenamefont{St-Onge, Sun, Allard,
  H{\'e}bert-Dufresne, and Bianconi}}]{st2021universal}
\bibinfo{author}{\bibfnamefont{G.}~\bibnamefont{St-Onge}},
  \bibinfo{author}{\bibfnamefont{H.}~\bibnamefont{Sun}},
  \bibinfo{author}{\bibfnamefont{A.}~\bibnamefont{Allard}},
  \bibinfo{author}{\bibfnamefont{L.}~\bibnamefont{H{\'e}bert-Dufresne}},
  \bibnamefont{and} \bibinfo{author}{\bibfnamefont{G.}~\bibnamefont{Bianconi}},
  \bibinfo{journal}{Physical Review Letters} \textbf{\bibinfo{volume}{127}},
  \bibinfo{pages}{158301} (\bibinfo{year}{2021}{\natexlab{a}}).

\bibitem[{\citenamefont{Li et~al.}(2021)\citenamefont{Li, Deng, Han,
  Alfaro-Bittner, Barzel, and Boccaletti}}]{li2021contagion}
\bibinfo{author}{\bibfnamefont{Z.}~\bibnamefont{Li}},
  \bibinfo{author}{\bibfnamefont{Z.}~\bibnamefont{Deng}},
  \bibinfo{author}{\bibfnamefont{Z.}~\bibnamefont{Han}},
  \bibinfo{author}{\bibfnamefont{K.}~\bibnamefont{Alfaro-Bittner}},
  \bibinfo{author}{\bibfnamefont{B.}~\bibnamefont{Barzel}}, \bibnamefont{and}
  \bibinfo{author}{\bibfnamefont{S.}~\bibnamefont{Boccaletti}},
  \bibinfo{journal}{Chaos, Solitons \& Fractals}
  \textbf{\bibinfo{volume}{152}}, \bibinfo{pages}{111307}
  (\bibinfo{year}{2021}).

\bibitem[{\citenamefont{St-Onge
  et~al.}(2021{\natexlab{b}})\citenamefont{St-Onge, Iacopini, Latora, Barrat,
  Petri, Allard, and H{\'e}bert-Dufresne}}]{st2021influential}
\bibinfo{author}{\bibfnamefont{G.}~\bibnamefont{St-Onge}},
  \bibinfo{author}{\bibfnamefont{I.}~\bibnamefont{Iacopini}},
  \bibinfo{author}{\bibfnamefont{V.}~\bibnamefont{Latora}},
  \bibinfo{author}{\bibfnamefont{A.}~\bibnamefont{Barrat}},
  \bibinfo{author}{\bibfnamefont{G.}~\bibnamefont{Petri}},
  \bibinfo{author}{\bibfnamefont{A.}~\bibnamefont{Allard}}, \bibnamefont{and}
  \bibinfo{author}{\bibfnamefont{L.}~\bibnamefont{H{\'e}bert-Dufresne}},
  \bibinfo{journal}{arXiv preprint arXiv:2105.07092}
  (\bibinfo{year}{2021}{\natexlab{b}}).

\bibitem[{\citenamefont{de~Arruda et~al.}(2021)\citenamefont{de~Arruda,
  Tizzani, and Moreno}}]{de2021phase}
\bibinfo{author}{\bibfnamefont{G.~F.} \bibnamefont{de~Arruda}},
  \bibinfo{author}{\bibfnamefont{M.}~\bibnamefont{Tizzani}}, \bibnamefont{and}
  \bibinfo{author}{\bibfnamefont{Y.}~\bibnamefont{Moreno}},
  \bibinfo{journal}{Communications Physics} \textbf{\bibinfo{volume}{4}},
  \bibinfo{pages}{1} (\bibinfo{year}{2021}).

\bibitem[{\citenamefont{Li et~al.}(2022)\citenamefont{Li, Xue, Pan, Lin, and
  Wang}}]{li2022competing}
\bibinfo{author}{\bibfnamefont{W.}~\bibnamefont{Li}},
  \bibinfo{author}{\bibfnamefont{X.}~\bibnamefont{Xue}},
  \bibinfo{author}{\bibfnamefont{L.}~\bibnamefont{Pan}},
  \bibinfo{author}{\bibfnamefont{T.}~\bibnamefont{Lin}}, \bibnamefont{and}
  \bibinfo{author}{\bibfnamefont{W.}~\bibnamefont{Wang}},
  \bibinfo{journal}{Applied Mathematics and Computation}
  \textbf{\bibinfo{volume}{412}}, \bibinfo{pages}{126595}
  (\bibinfo{year}{2022}).

\bibitem[{\citenamefont{Neuh{\"a}user et~al.}(2020)\citenamefont{Neuh{\"a}user,
  Mellor, and Lambiotte}}]{neuhauser2020multibody}
\bibinfo{author}{\bibfnamefont{L.}~\bibnamefont{Neuh{\"a}user}},
  \bibinfo{author}{\bibfnamefont{A.}~\bibnamefont{Mellor}}, \bibnamefont{and}
  \bibinfo{author}{\bibfnamefont{R.}~\bibnamefont{Lambiotte}},
  \bibinfo{journal}{Physical Review E} \textbf{\bibinfo{volume}{101}},
  \bibinfo{pages}{032310} (\bibinfo{year}{2020}).

\bibitem[{\citenamefont{Sahasrabuddhe et~al.}(2021)\citenamefont{Sahasrabuddhe,
  Neuh{\"a}user, and Lambiotte}}]{sahasrabuddhe2021modelling}
\bibinfo{author}{\bibfnamefont{R.}~\bibnamefont{Sahasrabuddhe}},
  \bibinfo{author}{\bibfnamefont{L.}~\bibnamefont{Neuh{\"a}user}},
  \bibnamefont{and}
  \bibinfo{author}{\bibfnamefont{R.}~\bibnamefont{Lambiotte}},
  \bibinfo{journal}{Journal of Physics: Complexity}
  \textbf{\bibinfo{volume}{2}}, \bibinfo{pages}{025006} (\bibinfo{year}{2021}).

\bibitem[{\citenamefont{Hickok et~al.}(2021)\citenamefont{Hickok, Kureh,
  Brooks, Feng, and Porter}}]{hickok2021bounded}
\bibinfo{author}{\bibfnamefont{A.}~\bibnamefont{Hickok}},
  \bibinfo{author}{\bibfnamefont{Y.}~\bibnamefont{Kureh}},
  \bibinfo{author}{\bibfnamefont{H.~Z.} \bibnamefont{Brooks}},
  \bibinfo{author}{\bibfnamefont{M.}~\bibnamefont{Feng}}, \bibnamefont{and}
  \bibinfo{author}{\bibfnamefont{M.~A.} \bibnamefont{Porter}},
  \bibinfo{journal}{arXiv preprint arXiv:2102.06825}  (\bibinfo{year}{2021}).

\bibitem[{\citenamefont{Ziegler et~al.}(2021)\citenamefont{Ziegler, Skardal,
  Dutta, and Taylor}}]{ziegler2021balanced}
\bibinfo{author}{\bibfnamefont{C.}~\bibnamefont{Ziegler}},
  \bibinfo{author}{\bibfnamefont{P.~S.} \bibnamefont{Skardal}},
  \bibinfo{author}{\bibfnamefont{H.}~\bibnamefont{Dutta}}, \bibnamefont{and}
  \bibinfo{author}{\bibfnamefont{D.}~\bibnamefont{Taylor}},
  \bibinfo{journal}{arXiv preprint arXiv:2112.01070}  (\bibinfo{year}{2021}).

\bibitem[{\citenamefont{Neuh{\"a}user et~al.}(2021)\citenamefont{Neuh{\"a}user,
  Lambiotte, and Schaub}}]{neuhauser2021consensus}
\bibinfo{author}{\bibfnamefont{L.}~\bibnamefont{Neuh{\"a}user}},
  \bibinfo{author}{\bibfnamefont{R.}~\bibnamefont{Lambiotte}},
  \bibnamefont{and} \bibinfo{author}{\bibfnamefont{M.~T.}
  \bibnamefont{Schaub}}, \bibinfo{journal}{arXiv preprint arXiv:2109.04985}
  (\bibinfo{year}{2021}).

\bibitem[{\citenamefont{Horstmeyer and Kuehn}(2020)}]{horstmeyer2020adaptive}
\bibinfo{author}{\bibfnamefont{L.}~\bibnamefont{Horstmeyer}} \bibnamefont{and}
  \bibinfo{author}{\bibfnamefont{C.}~\bibnamefont{Kuehn}},
  \bibinfo{journal}{Physical Review E} \textbf{\bibinfo{volume}{101}},
  \bibinfo{pages}{022305} (\bibinfo{year}{2020}).

\bibitem[{\citenamefont{DeVille}(2021)}]{deville2021consensus}
\bibinfo{author}{\bibfnamefont{L.}~\bibnamefont{DeVille}},
  \bibinfo{journal}{Chaos: An Interdisciplinary Journal of Nonlinear Science}
  \textbf{\bibinfo{volume}{31}}, \bibinfo{pages}{023137}
  (\bibinfo{year}{2021}).

\bibitem[{\citenamefont{Guo et~al.}(2021)\citenamefont{Guo, Jia,
  Sendi{\~n}a-Nadal, Zhang, Wang, Li, Alfaro-Bittner, Moreno, and
  Boccaletti}}]{guo2021evolutionary}
\bibinfo{author}{\bibfnamefont{H.}~\bibnamefont{Guo}},
  \bibinfo{author}{\bibfnamefont{D.}~\bibnamefont{Jia}},
  \bibinfo{author}{\bibfnamefont{I.}~\bibnamefont{Sendi{\~n}a-Nadal}},
  \bibinfo{author}{\bibfnamefont{M.}~\bibnamefont{Zhang}},
  \bibinfo{author}{\bibfnamefont{Z.}~\bibnamefont{Wang}},
  \bibinfo{author}{\bibfnamefont{X.}~\bibnamefont{Li}},
  \bibinfo{author}{\bibfnamefont{K.}~\bibnamefont{Alfaro-Bittner}},
  \bibinfo{author}{\bibfnamefont{Y.}~\bibnamefont{Moreno}}, \bibnamefont{and}
  \bibinfo{author}{\bibfnamefont{S.}~\bibnamefont{Boccaletti}},
  \bibinfo{journal}{Chaos, Solitons \& Fract} \textbf{\bibinfo{volume}{150}},
  \bibinfo{pages}{111103} (\bibinfo{year}{2021}).

\bibitem[{\citenamefont{Nowak}(2006)}]{nowak2006five}
\bibinfo{author}{\bibfnamefont{M.~A.} \bibnamefont{Nowak}},
  \bibinfo{journal}{Science} \textbf{\bibinfo{volume}{314}},
  \bibinfo{pages}{1560} (\bibinfo{year}{2006}).

\bibitem[{\citenamefont{Perc et~al.}(2017)\citenamefont{Perc, Jordan, Rand,
  Wang, Boccaletti, and Szolnoki}}]{perc2017statistical}
\bibinfo{author}{\bibfnamefont{M.}~\bibnamefont{Perc}},
  \bibinfo{author}{\bibfnamefont{J.~J.} \bibnamefont{Jordan}},
  \bibinfo{author}{\bibfnamefont{D.~G.} \bibnamefont{Rand}},
  \bibinfo{author}{\bibfnamefont{Z.}~\bibnamefont{Wang}},
  \bibinfo{author}{\bibfnamefont{S.}~\bibnamefont{Boccaletti}},
  \bibnamefont{and} \bibinfo{author}{\bibfnamefont{A.}~\bibnamefont{Szolnoki}},
  \bibinfo{journal}{Physics Reports} \textbf{\bibinfo{volume}{687}},
  \bibinfo{pages}{1} (\bibinfo{year}{2017}).

\bibitem[{\citenamefont{Perc and Szolnoki}(2010)}]{perc2010coevolutionary}
\bibinfo{author}{\bibfnamefont{M.}~\bibnamefont{Perc}} \bibnamefont{and}
  \bibinfo{author}{\bibfnamefont{A.}~\bibnamefont{Szolnoki}},
  \bibinfo{journal}{BioSystems} \textbf{\bibinfo{volume}{99}},
  \bibinfo{pages}{109} (\bibinfo{year}{2010}).

\bibitem[{\citenamefont{Szolnoki et~al.}(2014)\citenamefont{Szolnoki, Mobilia,
  Jiang, Szczesny, Rucklidge, and Perc}}]{szolnoki2014cyclic}
\bibinfo{author}{\bibfnamefont{A.}~\bibnamefont{Szolnoki}},
  \bibinfo{author}{\bibfnamefont{M.}~\bibnamefont{Mobilia}},
  \bibinfo{author}{\bibfnamefont{L.-L.} \bibnamefont{Jiang}},
  \bibinfo{author}{\bibfnamefont{B.}~\bibnamefont{Szczesny}},
  \bibinfo{author}{\bibfnamefont{A.~M.} \bibnamefont{Rucklidge}},
  \bibnamefont{and} \bibinfo{author}{\bibfnamefont{M.}~\bibnamefont{Perc}},
  \bibinfo{journal}{Journal of the Royal Society Interface}
  \textbf{\bibinfo{volume}{11}}, \bibinfo{pages}{20140735}
  (\bibinfo{year}{2014}).

\bibitem[{\citenamefont{Hilbe et~al.}(2014)\citenamefont{Hilbe, Wu, Traulsen,
  and Nowak}}]{hilbe2014cooperation}
\bibinfo{author}{\bibfnamefont{C.}~\bibnamefont{Hilbe}},
  \bibinfo{author}{\bibfnamefont{B.}~\bibnamefont{Wu}},
  \bibinfo{author}{\bibfnamefont{A.}~\bibnamefont{Traulsen}}, \bibnamefont{and}
  \bibinfo{author}{\bibfnamefont{M.~A.} \bibnamefont{Nowak}},
  \bibinfo{journal}{Proceedings of the National Academy of Sciences}
  \textbf{\bibinfo{volume}{111}}, \bibinfo{pages}{16425}
  (\bibinfo{year}{2014}).

\bibitem[{\citenamefont{Perc et~al.}(2013)\citenamefont{Perc,
  G{\'o}mez-Gardenes, Szolnoki, Flor{\'\i}a, and
  Moreno}}]{perc2013evolutionary}
\bibinfo{author}{\bibfnamefont{M.}~\bibnamefont{Perc}},
  \bibinfo{author}{\bibfnamefont{J.}~\bibnamefont{G{\'o}mez-Gardenes}},
  \bibinfo{author}{\bibfnamefont{A.}~\bibnamefont{Szolnoki}},
  \bibinfo{author}{\bibfnamefont{L.~M.} \bibnamefont{Flor{\'\i}a}},
  \bibnamefont{and} \bibinfo{author}{\bibfnamefont{Y.}~\bibnamefont{Moreno}},
  \bibinfo{journal}{Journal of the Royal Society Interface}
  \textbf{\bibinfo{volume}{10}}, \bibinfo{pages}{20120997}
  (\bibinfo{year}{2013}).

\bibitem[{\citenamefont{Perc and Szolnoki}(2008)}]{perc2008social}
\bibinfo{author}{\bibfnamefont{M.}~\bibnamefont{Perc}} \bibnamefont{and}
  \bibinfo{author}{\bibfnamefont{A.}~\bibnamefont{Szolnoki}},
  \bibinfo{journal}{Physical Review E} \textbf{\bibinfo{volume}{77}},
  \bibinfo{pages}{011904} (\bibinfo{year}{2008}).

\bibitem[{\citenamefont{Li et~al.}(2014)\citenamefont{Li, Wu, and
  Wang}}]{li2014cooperation}
\bibinfo{author}{\bibfnamefont{A.}~\bibnamefont{Li}},
  \bibinfo{author}{\bibfnamefont{B.}~\bibnamefont{Wu}}, \bibnamefont{and}
  \bibinfo{author}{\bibfnamefont{L.}~\bibnamefont{Wang}},
  \bibinfo{journal}{Scientific Reports} \textbf{\bibinfo{volume}{4}},
  \bibinfo{pages}{5536} (\bibinfo{year}{2014}).

\bibitem[{\citenamefont{Pe{\~n}a et~al.}(2016)\citenamefont{Pe{\~n}a, Wu,
  Arranz, and Traulsen}}]{pena2016evolutionary}
\bibinfo{author}{\bibfnamefont{J.}~\bibnamefont{Pe{\~n}a}},
  \bibinfo{author}{\bibfnamefont{B.}~\bibnamefont{Wu}},
  \bibinfo{author}{\bibfnamefont{J.}~\bibnamefont{Arranz}}, \bibnamefont{and}
  \bibinfo{author}{\bibfnamefont{A.}~\bibnamefont{Traulsen}},
  \bibinfo{journal}{PLoS Computational Biology} \textbf{\bibinfo{volume}{12}},
  \bibinfo{pages}{e1005059} (\bibinfo{year}{2016}).

\bibitem[{\citenamefont{Li et~al.}(2016)\citenamefont{Li, Broom, Du, and
  Wang}}]{li2016evolutionary}
\bibinfo{author}{\bibfnamefont{A.}~\bibnamefont{Li}},
  \bibinfo{author}{\bibfnamefont{M.}~\bibnamefont{Broom}},
  \bibinfo{author}{\bibfnamefont{J.}~\bibnamefont{Du}}, \bibnamefont{and}
  \bibinfo{author}{\bibfnamefont{L.}~\bibnamefont{Wang}},
  \bibinfo{journal}{Physical Review E} \textbf{\bibinfo{volume}{93}},
  \bibinfo{pages}{022407} (\bibinfo{year}{2016}).

\bibitem[{\citenamefont{Zhou et~al.}(2018)\citenamefont{Zhou, Wu, Vasconcelos,
  and Wang}}]{zhou2018simple}
\bibinfo{author}{\bibfnamefont{L.}~\bibnamefont{Zhou}},
  \bibinfo{author}{\bibfnamefont{B.}~\bibnamefont{Wu}},
  \bibinfo{author}{\bibfnamefont{V.~V.} \bibnamefont{Vasconcelos}},
  \bibnamefont{and} \bibinfo{author}{\bibfnamefont{L.}~\bibnamefont{Wang}},
  \bibinfo{journal}{Physical Review E} \textbf{\bibinfo{volume}{98}},
  \bibinfo{pages}{062124} (\bibinfo{year}{2018}).

\bibitem[{\citenamefont{Su et~al.}(2019{\natexlab{a}})\citenamefont{Su, Zhou,
  and Wang}}]{su2019evolutionary}
\bibinfo{author}{\bibfnamefont{Q.}~\bibnamefont{Su}},
  \bibinfo{author}{\bibfnamefont{L.}~\bibnamefont{Zhou}}, \bibnamefont{and}
  \bibinfo{author}{\bibfnamefont{L.}~\bibnamefont{Wang}},
  \bibinfo{journal}{PLoS Computational Biology} \textbf{\bibinfo{volume}{15}},
  \bibinfo{pages}{e1006947} (\bibinfo{year}{2019}{\natexlab{a}}).

\bibitem[{\citenamefont{Su et~al.}(2018)\citenamefont{Su, Wang, and
  Stanley}}]{su2018understanding}
\bibinfo{author}{\bibfnamefont{Q.}~\bibnamefont{Su}},
  \bibinfo{author}{\bibfnamefont{L.}~\bibnamefont{Wang}}, \bibnamefont{and}
  \bibinfo{author}{\bibfnamefont{H.~E.} \bibnamefont{Stanley}},
  \bibinfo{journal}{New Journal of Physics} \textbf{\bibinfo{volume}{20}},
  \bibinfo{pages}{103030} (\bibinfo{year}{2018}).

\bibitem[{\citenamefont{Su et~al.}(2019{\natexlab{b}})\citenamefont{Su, Li,
  Wang, and Eugene~Stanley}}]{su2019spatial}
\bibinfo{author}{\bibfnamefont{Q.}~\bibnamefont{Su}},
  \bibinfo{author}{\bibfnamefont{A.}~\bibnamefont{Li}},
  \bibinfo{author}{\bibfnamefont{L.}~\bibnamefont{Wang}}, \bibnamefont{and}
  \bibinfo{author}{\bibfnamefont{H.}~\bibnamefont{Eugene~Stanley}},
  \bibinfo{journal}{Proceedings of the Royal Society B}
  \textbf{\bibinfo{volume}{286}}, \bibinfo{pages}{20190041}
  (\bibinfo{year}{2019}{\natexlab{b}}).

\bibitem[{\citenamefont{Burgio et~al.}(2020)\citenamefont{Burgio, Matamalas,
  G{\'o}mez, and Arenas}}]{burgio2020evolution}
\bibinfo{author}{\bibfnamefont{G.}~\bibnamefont{Burgio}},
  \bibinfo{author}{\bibfnamefont{J.~T.} \bibnamefont{Matamalas}},
  \bibinfo{author}{\bibfnamefont{S.}~\bibnamefont{G{\'o}mez}},
  \bibnamefont{and} \bibinfo{author}{\bibfnamefont{A.}~\bibnamefont{Arenas}},
  \bibinfo{journal}{Entropy} \textbf{\bibinfo{volume}{22}},
  \bibinfo{pages}{744} (\bibinfo{year}{2020}).

\bibitem[{\citenamefont{Alvarez-Rodriguez
  et~al.}(2021)\citenamefont{Alvarez-Rodriguez, Battiston, de~Arruda, Moreno,
  Perc, and Latora}}]{alvarez2021evolutionary}
\bibinfo{author}{\bibfnamefont{U.}~\bibnamefont{Alvarez-Rodriguez}},
  \bibinfo{author}{\bibfnamefont{F.}~\bibnamefont{Battiston}},
  \bibinfo{author}{\bibfnamefont{G.~F.} \bibnamefont{de~Arruda}},
  \bibinfo{author}{\bibfnamefont{Y.}~\bibnamefont{Moreno}},
  \bibinfo{author}{\bibfnamefont{M.}~\bibnamefont{Perc}}, \bibnamefont{and}
  \bibinfo{author}{\bibfnamefont{V.}~\bibnamefont{Latora}},
  \bibinfo{journal}{Nature Human Behaviour} \textbf{\bibinfo{volume}{5}},
  \bibinfo{pages}{586} (\bibinfo{year}{2021}).

\bibitem[{\citenamefont{Schlager et~al.}(2021)\citenamefont{Schlager,
  Clau{\ss}, and Kuehn}}]{schlager2021stability}
\bibinfo{author}{\bibfnamefont{D.}~\bibnamefont{Schlager}},
  \bibinfo{author}{\bibfnamefont{K.}~\bibnamefont{Clau{\ss}}},
  \bibnamefont{and} \bibinfo{author}{\bibfnamefont{C.}~\bibnamefont{Kuehn}},
  \bibinfo{journal}{arXiv preprint arXiv:2111.07852}  (\bibinfo{year}{2021}).

\bibitem[{\citenamefont{Civilini et~al.}(2021)\citenamefont{Civilini, Anbarci,
  and Latora}}]{civilini2021evolutionary}
\bibinfo{author}{\bibfnamefont{A.}~\bibnamefont{Civilini}},
  \bibinfo{author}{\bibfnamefont{N.}~\bibnamefont{Anbarci}}, \bibnamefont{and}
  \bibinfo{author}{\bibfnamefont{V.}~\bibnamefont{Latora}},
  \bibinfo{journal}{Physical Review Letters} \textbf{\bibinfo{volume}{127}},
  \bibinfo{pages}{268301} (\bibinfo{year}{2021}).

\bibitem[{\citenamefont{Kumar et~al.}(2021)\citenamefont{Kumar, Chowdhary,
  Capraro, and Perc}}]{kumar2021evolution}
\bibinfo{author}{\bibfnamefont{A.}~\bibnamefont{Kumar}},
  \bibinfo{author}{\bibfnamefont{S.}~\bibnamefont{Chowdhary}},
  \bibinfo{author}{\bibfnamefont{V.}~\bibnamefont{Capraro}}, \bibnamefont{and}
  \bibinfo{author}{\bibfnamefont{M.}~\bibnamefont{Perc}},
  \bibinfo{journal}{Physical Review E} \textbf{\bibinfo{volume}{104}},
  \bibinfo{pages}{054308} (\bibinfo{year}{2021}).

\bibitem[{\citenamefont{Mukherjee and Steenbergen}(2016)}]{mukherjee2016random}
\bibinfo{author}{\bibfnamefont{S.}~\bibnamefont{Mukherjee}} \bibnamefont{and}
  \bibinfo{author}{\bibfnamefont{J.}~\bibnamefont{Steenbergen}},
  \bibinfo{journal}{Random structures \& Algorithms}
  \textbf{\bibinfo{volume}{49}}, \bibinfo{pages}{379} (\bibinfo{year}{2016}).

\bibitem[{\citenamefont{Parzanchevski and
  Rosenthal}(2017)}]{parzanchevski2017simplicial}
\bibinfo{author}{\bibfnamefont{O.}~\bibnamefont{Parzanchevski}}
  \bibnamefont{and}
  \bibinfo{author}{\bibfnamefont{R.}~\bibnamefont{Rosenthal}},
  \bibinfo{journal}{Random Structures \& Algorithms}
  \textbf{\bibinfo{volume}{50}}, \bibinfo{pages}{225} (\bibinfo{year}{2017}).

\bibitem[{\citenamefont{Carletti
  et~al.}(2020{\natexlab{b}})\citenamefont{Carletti, Battiston, Cencetti, and
  Fanelli}}]{carletti2020random}
\bibinfo{author}{\bibfnamefont{T.}~\bibnamefont{Carletti}},
  \bibinfo{author}{\bibfnamefont{F.}~\bibnamefont{Battiston}},
  \bibinfo{author}{\bibfnamefont{G.}~\bibnamefont{Cencetti}}, \bibnamefont{and}
  \bibinfo{author}{\bibfnamefont{D.}~\bibnamefont{Fanelli}},
  \bibinfo{journal}{Physical Review E} \textbf{\bibinfo{volume}{101}},
  \bibinfo{pages}{022308} (\bibinfo{year}{2020}{\natexlab{b}}).

\bibitem[{\citenamefont{Schaub et~al.}(2020)\citenamefont{Schaub, Benson, Horn,
  Lippner, and Jadbabaie}}]{schaub2020random}
\bibinfo{author}{\bibfnamefont{M.~T.} \bibnamefont{Schaub}},
  \bibinfo{author}{\bibfnamefont{A.~R.} \bibnamefont{Benson}},
  \bibinfo{author}{\bibfnamefont{P.}~\bibnamefont{Horn}},
  \bibinfo{author}{\bibfnamefont{G.}~\bibnamefont{Lippner}}, \bibnamefont{and}
  \bibinfo{author}{\bibfnamefont{A.}~\bibnamefont{Jadbabaie}},
  \bibinfo{journal}{SIAM Review} \textbf{\bibinfo{volume}{62}},
  \bibinfo{pages}{353} (\bibinfo{year}{2020}).

\bibitem[{\citenamefont{Carletti et~al.}(2021)\citenamefont{Carletti, Fanelli,
  and Lambiotte}}]{carletti2021random}
\bibinfo{author}{\bibfnamefont{T.}~\bibnamefont{Carletti}},
  \bibinfo{author}{\bibfnamefont{D.}~\bibnamefont{Fanelli}}, \bibnamefont{and}
  \bibinfo{author}{\bibfnamefont{R.}~\bibnamefont{Lambiotte}},
  \bibinfo{journal}{Journal of Physics: Complexity}
  \textbf{\bibinfo{volume}{2}}, \bibinfo{pages}{015011} (\bibinfo{year}{2021}).

\bibitem[{\citenamefont{Delvenne et~al.}(2010)\citenamefont{Delvenne, Yaliraki,
  and Barahona}}]{delvenne2010stability}
\bibinfo{author}{\bibfnamefont{J.-C.} \bibnamefont{Delvenne}},
  \bibinfo{author}{\bibfnamefont{S.~N.} \bibnamefont{Yaliraki}},
  \bibnamefont{and} \bibinfo{author}{\bibfnamefont{M.}~\bibnamefont{Barahona}},
  \bibinfo{journal}{Proceedings of the National Academy of Sciences}
  \textbf{\bibinfo{volume}{107}}, \bibinfo{pages}{12755}
  (\bibinfo{year}{2010}).

\bibitem[{\citenamefont{Torres and Bianconi}(2020)}]{torres2020simplicial}
\bibinfo{author}{\bibfnamefont{J.~J.} \bibnamefont{Torres}} \bibnamefont{and}
  \bibinfo{author}{\bibfnamefont{G.}~\bibnamefont{Bianconi}},
  \bibinfo{journal}{Journal of Physics: Complexity}
  \textbf{\bibinfo{volume}{1}}, \bibinfo{pages}{015002} (\bibinfo{year}{2020}).

\bibitem[{\citenamefont{Mill{\'a}n et~al.}(2021)\citenamefont{Mill{\'a}n,
  Ghorbanchian, Defenu, Battiston, and Bianconi}}]{millan2021local}
\bibinfo{author}{\bibfnamefont{A.~P.} \bibnamefont{Mill{\'a}n}},
  \bibinfo{author}{\bibfnamefont{R.}~\bibnamefont{Ghorbanchian}},
  \bibinfo{author}{\bibfnamefont{N.}~\bibnamefont{Defenu}},
  \bibinfo{author}{\bibfnamefont{F.}~\bibnamefont{Battiston}},
  \bibnamefont{and} \bibinfo{author}{\bibfnamefont{G.}~\bibnamefont{Bianconi}},
  \bibinfo{journal}{arXiv preprint arXiv:2102.12885}  (\bibinfo{year}{2021}).

\bibitem[{\citenamefont{Eriksson et~al.}(2021)\citenamefont{Eriksson, Edler,
  Rojas, de~Domenico, and Rosvall}}]{eriksson2021choosing}
\bibinfo{author}{\bibfnamefont{A.}~\bibnamefont{Eriksson}},
  \bibinfo{author}{\bibfnamefont{D.}~\bibnamefont{Edler}},
  \bibinfo{author}{\bibfnamefont{A.}~\bibnamefont{Rojas}},
  \bibinfo{author}{\bibfnamefont{M.}~\bibnamefont{de~Domenico}},
  \bibnamefont{and} \bibinfo{author}{\bibfnamefont{M.}~\bibnamefont{Rosvall}},
  \bibinfo{journal}{Communications Physics} \textbf{\bibinfo{volume}{4}},
  \bibinfo{pages}{1} (\bibinfo{year}{2021}).

\bibitem[{\citenamefont{Majhi et~al.}(2019{\natexlab{b}})\citenamefont{Majhi,
  Bera, Ghosh, and Perc}}]{majhi2019chimera}
\bibinfo{author}{\bibfnamefont{S.}~\bibnamefont{Majhi}},
  \bibinfo{author}{\bibfnamefont{B.~K.} \bibnamefont{Bera}},
  \bibinfo{author}{\bibfnamefont{D.}~\bibnamefont{Ghosh}}, \bibnamefont{and}
  \bibinfo{author}{\bibfnamefont{M.}~\bibnamefont{Perc}},
  \bibinfo{journal}{Physics of Life Reviews} \textbf{\bibinfo{volume}{28}},
  \bibinfo{pages}{100} (\bibinfo{year}{2019}{\natexlab{b}}).

\bibitem[{\citenamefont{Gross and Blasius}(2008)}]{gross2008adaptive}
\bibinfo{author}{\bibfnamefont{T.}~\bibnamefont{Gross}} \bibnamefont{and}
  \bibinfo{author}{\bibfnamefont{B.}~\bibnamefont{Blasius}},
  \bibinfo{journal}{Journal of the Royal Society Interface}
  \textbf{\bibinfo{volume}{5}}, \bibinfo{pages}{259} (\bibinfo{year}{2008}).

\end{thebibliography}
\end{document}